%% file: main.tex
\documentclass[journal]{IEEEtran}

\IEEEoverridecommandlockouts                              
\usepackage{lastpage}
\usepackage{fancyhdr}

\usepackage[pdftex]{graphicx}
\graphicspath{{figure/}}
\usepackage{amsmath}
\usepackage{amssymb}
\usepackage{url}

\usepackage{amsthm} 
\usepackage{cite}
\usepackage{bm} 

\usepackage{algorithm}
\usepackage{algorithmic}

\usepackage{amssymb}
\usepackage{color}

\newtheorem{remark}{\textbf{Remark}}

\newcommand{\mc}{\mathcal}
\newcommand{\mb}{\mathbf}

\newcommand{\R}{\mathbb{R}}
\newcommand{\bx}{\bm{x}}
\newcommand{\by}{\bm{y}}

\newcommand{\bxi}{\bm{\xi}}
\newcommand{\bga}{\bm{\gamma}}
\newcommand{\bmu}{\bm{\mu}}

\newcommand{\bp}{\bm{p}}
\newcommand{\bq}{\bm{q}}
\newcommand{\bv}{\bm{v}}

\usepackage{epstopdf}

\newboolean{showcomments}
\setboolean{showcomments}{true}

\usepackage{multirow}

\allowdisplaybreaks

\begin{document}

\input{title}
	
	\input{nomenclature}

\input{introduction}

	\input{system_model}

\input{aggregation_method}

	\input{Solution_alg}

\input{case_study}

	\input{conclusion}

\input{appendix}

\input{reference}\input{biography}
	
\end{document}

%% file: title.tex
\title{Leveraging Two-Stage Adaptive Robust Optimization for Power  Flexibility Aggregation
}

\author{Xin Chen,~\IEEEmembership{Student Member,~IEEE,}
 Na Li,~\IEEEmembership{Member,~IEEE}
	\thanks{  X. Chen and N. Li are with the School of Engineering and Applied Sciences, Harvard University, USA. Email: (chen\_xin@g.harvard.edu, nali@seas.harvard.edu). 
	}
	\thanks{ 
The work was supported by
NSF 1608509, NSF CAREER 1553407, AFOSR YIP, and Harvard Climate Change Solution Funds.} 
}

\maketitle
%\thispagestyle{empty}  % plain
%\pagestyle{empty}      % plain for page number

% -----------------------------------------------

\begin{abstract}

Adaptive robust optimization (ARO) is a well-known technique to deal with the parameter uncertainty in optimization problems. While the ARO framework can actually be borrowed to solve some special problems without uncertain parameters, such as the power flexibility aggregation problem studied in this paper.
To effectively harness the significant flexibility from massive distributed energy resources (DERs), power flexibility aggregation is performed for a distribution system to compute  the feasible region of the exchanged power at the substation over  time.
Based on  two-stage ARO, 
this paper proposes a novel method  to aggregate  system-level multi-period
 power flexibility, considering heterogeneous DER facilities, network operational constraints, and an unbalanced power flow model. This method is applicable to aggregate only the active (or reactive) power, and the joint active-reactive power domain. Accordingly,  two  power  aggregation models with two-stage optimization are developed: one 
 focuses on aggregating active power and
 computes its optimal feasible intervals over multiple periods, and the other solves the optimal elliptical feasible regions for the aggregate active-reactive power.
 By leveraging the ARO technique,  the disaggregation feasibility of the obtained feasible regions is guaranteed with optimality. Numerical simulations  on a real-world distribution feeder with 126 multi-phase nodes demonstrate the effectiveness of the proposed method.
\end{abstract}


%% file: nomenclature.tex
	\section*{Nomenclature}

\addcontentsline{toc}{section}{Nomenclature}

\subsection{Parameters}
\begin{IEEEdescription}[\IEEEusemathlabelsep\IEEEsetlabelwidth{$V_1,V_2,V_3$}]
%	\item [$N,\, L$] Number of buses (except the substation bus), number of distribution lines.
%	\item [$N_Y,\,N_\Delta$] Number of buses with wye-, delta-connection.
%	\item [$L$] Number of distribution lines.
	\item [$\bar{\bm{v}},\,\underline{\bm{v}}$]  Upper, lower limits of the three-phase nodal voltage magnitudes for all buses. 
	\item [$\bar{\bm{i}},\, \underline{\bm{i}} $] Upper,  lower limits of the three-phase line current magnitudes for all distribution lines.
	\item [$\bar{p}_{i,t}^{\mathrm{pv},\psi},\underline{p}_{i,t}^{\mathrm{pv},\psi}$] Upper, lower 
	limits of active PV power generation in phase $\psi$ of bus $i$ at time $t$.
	\item [$\bar{s}_{i,t}^{\mathrm{pv},\psi}$] 
    Apparent power capacity of PV units in phase $\psi$ of bus $i$ at time $t$.
    \item [$\bar{p}_{i,t}^{\mathrm{es},\psi},\underline{p}_{i,t}^{\mathrm{es},\psi}$]  Upper, lower 
    limits of active power output of ES devices in phase $\psi$ of bus $i$ at time $t$.
    \item [$\bar{s}_{i,t}^{\mathrm{es},\psi}$] Apparent power capacity of ES devices in phase $\psi$ of bus $i$ at time $t$.
    \item [$\bar{E}^{\mathrm{es}}_i, \underline{E}^{\mathrm{es}}_i$] 
     Upper, lower limits for state of charge of  ES devices at bus $i$.
     \item [$\bar{p}_{i,t}^{\mathrm{d},\psi},\underline{p}_{i,t}^{\mathrm{d},\psi}$]  Upper, lower limits for  controllable active loads in phase $\psi$ of bus $i$ at time $t$.
  \item  [$\bar{E}_i^{\mathrm{d}}, \underline{E}_i^{\mathrm{d}}$] Maximum, minimum total energy  required to consume for controllable loads at bus $i$.
     \item [$F_{i,t}^{\mathrm{out}}$] Outside temperature for HVAC systems
      at bus $i$ at time $t$.
     \item [$\bar{F}_{i},\underline{F}_{i}$] Upper, lower limits of comfortable temperature zone for HVAC systems
     at bus $i$.
     \item [$\Delta t$]  Length of each time slot under discretized time horizon.
     % Time granularity between two adjacent time slots.
   %  \item [$ P_{i,t}^{L,\psi}, Q_{i,t}^{L,\psi}$] Active, reactive uncontrollable loads in phase $\psi$ of bus $i$ at time $t$.  
\end{IEEEdescription}	

\subsection{Variables}
	\begin{IEEEdescription}[\IEEEusemathlabelsep\IEEEsetlabelwidth{$V_1,V_2,V_3$}]
%		\item[$\mb{p}_Y,\mb{q}_Y$\\$\in \mathbb{R}^{3N_Y}$] Column vector of the three-phase active, reactive power injection at the buses via wye-connection.
%		\item[$\mb{p}_\Delta,\mb{q}_\Delta$] Column vector of the three-phase active, reactive power injection at the buses via  delta-connection.
%\item[\smash{\begin{IEEEeqnarraybox*}[][t]{l}		 \bm{p}_Y, \bm{q}_Y\\	 	\hphantom{}\in \mathbb{R}^{3N_Y}		\end{IEEEeqnarraybox*}}] Column vector of the three-phase   active,		{reactive power injection  via  wye-connection.}
 %       \item[\smash{\begin{IEEEeqnarraybox*}[][t]{l}		\bm{p}_\Delta, \bm{q}_\Delta\\		\hphantom{}\in \mathbb{R}^{3N_\Delta}         \end{IEEEeqnarraybox*}}] Column vector of the three-phase   active,    {reactive  power injection  via  delta-connection.}
%        \item[$\bx$]  $:=\left[\mb{p}_Y^\top,\mb{q}_Y^\top,\mb{p}_\Delta^\top,\mb{q}_\Delta^\top \right]^\top$, column vector collecting all three-phase  power injection.
        	\item[$\bm{v}_t\!\in\mathbb{R}^{3|\mathcal{N}|}$] Column vector collecting the three-phase nodal voltage magnitudes for all buses at time $t$.
        \item[$\bm{i}_{t}\in \mathbb{R}^{3|\mathcal{E}|}$] Column vector collecting the three-phase  line current magnitudes for all lines at time $t$.
        \item[$\bm{p}_{0},\!\bm{q}_{0}\!\in\!\mathbb{R}^{T}$] Column vector of the total three-phase net active, reactive  power injection at the substation. 
         \item [$p_{0,t},q_{0,t}$] Total net active, reactive power injection at the substation at time $t$.
       \item [$p_{i,t}^{\mathrm{pv},\psi},q_{i,t}^{\mathrm{pv},\psi}$] Active, reactive PV  power generation in phase $\psi$ of bus $i$ at time $t$.
       \item [$p_{i,t}^{\mathrm{es},\psi},q_{i,t}^{\mathrm{es},\psi}$] Active, reactive power output of ES devices in phase $\psi$ of bus $i$ at time $t$.
    \item  [$p_{i,t}^{\mathrm{cha},\psi}\!,\!p_{i,t}^{\mathrm{dis},\psi}$]
       Active  charging, discharging power of ES devices in phase $\psi$ of bus $i$ at time $t$. 
       \item [$E^{\mathrm{es}}_{i,t}$] State of charge of ES devices at bus $i$ at time $t$.
       \item [$p_{i,t}^{\mathrm{d},\psi},q_{i,t}^{\mathrm{d},\psi}$] Active, reactive controlled loads in phase $\psi$ of bus $i$ at time $t$.
       \item [$p_{i,t}^{\mathrm{hv},\psi},q_{i,t}^{\mathrm{hv},\psi}$]  Active, reactive HVAC loads in phase $\psi$ of bus $i$ at time $t$.
       \item [$F_{i,t}^{\mathrm{in}}$] Indoor temperature for HVAC systems
       at bus $i$ at time $t$.
\end{IEEEdescription}
\textbf{Note}: the same notations without  superscript $\psi$ denote the corresponding summation over phases, e.g. $p_{i,t}^{\mathrm{es}}: = \sum_{\psi}p_{i,t}^{\mathrm{es},\psi} $.

\subsection{Notations}
\begin{IEEEdescription}[\IEEEusemathlabelsep\IEEEsetlabelwidth{$\jmath:=\sqrt{-1}$}]
	\item[$\jmath$] $:=\sqrt{-1}$, i.e., the imaginary unit.
%		\item[$\mathbf{1}_n$] $n\times1$ column vector with all ones. 	
		\item[$|\cdot|$] Entry-wise absolute value of a vector or the cardinality of a set.
	\item[$||\bm{x}||$] $L$-2 norm of vector $\bx$.
	\item[$\mathrm{diag}(\bm{x})$]  Diagonal matrix with the elements of vector $\bm{x}$ in its diagonal.
	\item[$\bx\circ \by$] Entry-wise multiplication of vectors $\bx$ and $\by$.
	\item[$(\cdot)^\top,(\cdot)^{-1}$] Matrix transposition and matrix inverse.
	\item [$\bm{1}$] Column vector with all entries being one.
%	\item[$\frac{\partial f(x)}{\partial x}$] Partial derivative. 
\end{IEEEdescription}

%% file: introduction.tex
\section{Introduction}

%\IEEEPARstart{W}{ith} deepening penetration of renewable generation, such as wind power and solar energy, power systems  confront  emerging challenges in maintaining power balance due to inadequate reserve capacity. Thus additional sources of power flexibility are necessitated to  warrant secure and efficient system operation.

\IEEEPARstart{T}{he} recent decade has witnessed  a rapid proliferation of distributed energy resources (DERs) in distribution systems, 
 including dispatchable photovoltaic (PV) units, wind generators,  energy storage (ES) devices,  controllable loads,  etc.
 The coordinated dispatch of ubiquitous DERs is envisioned to release significant power flexibility and enable active interactions between transmission systems and distribution systems \cite{td-inter}. 
 However, managing a large population of DERs for system-wide optimization  and  control in real-time  is  challenging  due  to   high computational complexity. A promising solution to harness collective DER flexibility for transmission-distribution interaction is \emph{power flexibility aggregation}.
Specifically, 
 the power flexibility refers to the capability of a distribution system\footnote{The concept of power flexibility aggregation is applicable to other power sub-grids, such as micro-grids and regional grids, which interface with the bulk power system at the point of common coupling (PCC).}  to modulate its exchanged power (i.e., aggregate power) with the transmission system at the substation interface, and power flexibility aggregation is to characterize  the time-varying feasible region of the aggregate power.
This feasible region is essentially determined  by the internal DER operational conditions and the  network constraints. 
In this way,
 the complex configurations and states of a distribution feeder with massive DERs are represented in a concise and compact form. Then
 the transmission-distribution interaction can be achieved through
a hierarchical coordination framework, where each distribution feeder  acts as  a virtual power plant \cite{vpp} 
and reports its own aggregate feasible region.

%  In  \cite{tdc1,tdc2,tdc3}, decomposition  techniques, such as master-slave-splitting method, Benders decomposition, and dual decomposition, are employed to co-optimize transmission and distribution  in economic dispatch, reactive power optimization, and market clearing, which require frequent boundary information exchanges under a number of iterations.

With massive DER facilities and multiple time periods, procuring the exact feasible region of the aggregate power is computationally  impractical \cite{exact}, thus  most researches focus on approximation approaches.
References \cite{tcl1,tcl2,zono} use the polytope set   to describe the admissible power profile of an individual flexible load, 
then the aggregate flexibility is computed as the Minkowski sum of all these polytope sets.
%Reference \cite{zono} further proposes to use zonotope, a subclass of polytope, to depict the feasible  region of DER power, which leads to an efficient Minkowski summation.
In  \cite{proj}, the aggregate flexibility of heterogeneous deferrable loads is computed via  polytopic projection. References 
 \cite{ro2} builds a single-stage robust model to  schedule the  reserve capacities of the heating, ventilation and  air-conditioning (HVAC) systems%,considering the uncertainty of  forecast errors and power tracking signals
 . These researches  focus on a single type of DER, which may not be able to handle a variety of DERs with heterogeneous operational conditions. Moreover,  
the underlying 
 networks and  power flow  constraints are not taken into account.

%Most  studies above  only focus on a single type of DER, which may not be able to handle a variety of DERs with heterogeneous operational conditions. Moreover,  the underlying distribution networks and the associated power flow  constraints are not taken into account.  

For system-level flexibility aggregation, Monte-Carlo simulation based methods are proposed in \cite{ellip, MCS} to estimate the flexibility range of aggregate active and reactive
power (P-Q) with a large number of  sampling scenarios.
Reference \cite{rcc} computes the robust P-Q capability curve of a virtual power plant considering uncertain DER outputs and demand forecast errors. While these approaches only account for the aggregate flexibility at a single time snapshot, which may not capture the time-coupling  operational  constraints of inter-temporal DERs, such as energy storage devices and HVAC systems.
Reference \cite{opf1}
models  the multi-period feasible P-Q region with time-varying ellipsoids, and computes the corresponding parameters by a data-driven system identification procedure.  However, this method can not guarantee  \emph{disaggregation feasibility}, i.e., any  aggregate power trajectory within the obtained feasible region can be realized by appropriately dispatching DERs without violating  network or DER operational constraints. In \cite{our1}, some heuristic constraints associated with the upper and lower power trajectories  are added  to ensure disaggregation feasibility, but these constraints are conservative and  lead to sub-optimal aggregation solutions.
Therefore, it remains an unsolved task to develop  system-level multi-period power flexibility aggregation methods that incorporate both aggregation optimality and disaggregation feasibility.

Robust optimization (RO) \cite{roge} is a well-known technique for dealing with parameter uncertainty in optimization problems, and adaptive RO (ARO) \cite{aro} is further developed to reduce the conservativeness for the uncertain problems containing adaptive variables. Despite these facts,
interestingly,
 the ARO framework can be borrowed to address some special problems \emph{without uncertain parameters}. And we realize that the 
 power flexibility aggregation  studied in this paper is one of such special problems. 
 The intuition is that 
  the feasible region (of the aggregate power) is analogous to the uncertainty set (of the uncertain variables) in ARO. The requirement of disaggregation feasibility is exactly interpreted as the adaptive robust constraint, i.e., there must exist an associated feasible solution for any realization of the uncertain variables.
 Thus  power flexibility aggregation  can be formulated 
   as a two-stage ARO problem, where the first stage solves for the optimal feasible region of the aggregate power and the second stage guarantees the disaggregation feasibility. See Section \ref{sec:paggaro} for detailed explanations.

  The two-stage ARO method has been widely used in power system applications, such as  reactive power optimization \cite{rop1}, DER capacity assessment \cite{rop0,ropp1,ropp2},
  unit commitment \cite{rop2},  service restoration \cite{rop3}, etc., to tackle the uncertainty of renewable generation and load demand. In these works, the uncertainty set is predefined to restrain the uncertain variables,
  and ARO is used to obtain robust solutions that are immune to any scenarios within this uncertainty set. Distinguished from these work,  
  our idea lies in an innovative usage of ARO, where no parameter uncertainty is considered\footnote{We are aware that there exist uncertainty issues, such as uncertain load and renewable generation, in the power flexibility aggregation problem. This paper focuses on the deterministic aggregation methods, while the standard ARO can be further  applied to the proposed models to address
  the parameter uncertainty issues.} and the uncertainty set is not given but treated as the decision variable. The advantage of using ARO for power aggregation  is that  it provides theoretical guarantees on disaggregation feasibility with aggregation optimality. 
     Besides, mature ARO solution methods, e.g. the column-constraint-generation (CCG) algorithm \cite{ccg}, can be directly applied for efficient solution as well.

In this paper,  we study the system-level multi-period power flexibility aggregation for a distribution system, where a  variety of DER facilities,  a multi-phase unbalanced power flow model, and the network operational constraints are considered. In
particular, we characterize the  admissible aggregate power over time by a parameterized  set, and two-stage ARO models are established to solve for the maximum-volume parameterized  set inscribed inside the exact feasible region.  The main contributions of this paper are summarized as follows:

1) We propose a novel  system-level power  flexibility aggregation methodology by leveraging the  two-stage ARO technique, which can guarantee both aggregation optimality and disaggregation feasibility. Besides, the proposed method can be generalized as the framework to solve high-dimensional projection problems via ARO.

%propose a novel methodology for  system-level power flexibility aggregation by leveraging the concept of two-stage

2) Two concrete power flexibility aggregation models with two-stage optimization are developed for practical application: 
one (i.e., model (\ref{eq:APA})) solves the optimal feasible intervals for the aggregate active power over time, and the other (i.e.,  model (\ref{eq:ARPA}))  solves the optimal elliptical feasible regions for the time-variant aggregate P-Q domain.

The remainder of this paper is organized as follows: Section II introduces the multi-phase network and DER models. Section III presents the two-stage  power aggregation  method. Section IV 
develops the solution algorithm. Numerical tests are performed on a real distribution feeder in Section V, and conclusions are drawn in Section VI.

%% file: system_model.tex
% ------------------------------------------
\section{Distributed Energy Resource and Network Models}

Consider a multi-phase unbalanced distribution network described by the graph
$G({\mc{N}}_0,\mc{E})$, where $\mc{N}_0$ denotes the set of buses and $\mc{E}\subset \mc{N}_0\times \mc{N}_0$ denotes the set of  distribution lines connecting the buses.
Let 
 $\mc{N}_0:= \{0\} \cup \mc{N} $ with $\mc{N}:=\{1,2,\dots,N\}$ and bus $0$ denotes the  substation interface that exchanges power with the transmission system. 
Each electric device can be multi-phase wye-connected or delta-connected to the network \cite{wye}. 
 Denote  $\phi_{Y}:=\left\{a,b,c\right\}$ and $\phi_{\Delta}:=\left\{ab,bc,ca\right\}$. Then we use notation $\psi$ to describe the concrete connection manner of an electric device  with either $\psi\subseteq \phi_{Y}$ or $\psi\subseteq \phi_{\Delta}$. 
For instance,  $\psi=\left\{a\right\}$ if the device is wye-connected in phase A and only has the complex power injection $s^a := p^a + \jmath\, q^a$, while $\psi=\left\{ab,bc\right\}$ if it is delta-connected in phase AB and BC with the complex power injection $s^{ab}:= p^{ab} + \jmath\, q^{ab}$ and $s^{bc}:=p^{bc} + \jmath\, q^{bc}$.

%\begin{figure}[thpb]
%	\centering
%	\includegraphics[scale=0.795]{dw.pdf}
%	\caption{Illustration of wye-connection and delta-connection.}
%	\label{w_d}
%\end{figure}

\subsection{Distributed Energy Resource Model}
%{\color{red} if the models are very similar to our previous paper, you can try to refer to the other paper to reduce the overlap or move it to appendix. You can keep them here too but please make sure the overlap is minimal.}

Given a discrete-time horizon $\mc{T}:=\{1,2,\cdots,T\}$, we consider several typical DERs, including dispatchable PV units, ES devices, directly controllable loads, and HVAC systems. 
 The associated DER operational models are established as follows, where $\mathcal{N}_{\{\mathrm{pv},\, \mathrm{es},\, \mathrm{dcl},\, \mathrm{hv}\}}$ denotes the set of buses connected with the corresponding DER devices. 

\subsubsection{Dispatchable PV Units}
$\forall i \in \mc{N}_{\mathrm{pv}}, \ t\in \mc{T}$
\begin{subequations} \label{pv}
	\begin{gather} 
	\underline{p}_{i,t}^{\mathrm{pv},\psi} \leq  p_{i,t}^{\mathrm{pv},\psi} \leq \bar{p}_{i,t}^{\mathrm{pv},\psi} \\
	(p_{i,t}^{\mathrm{pv},\psi})^2+(q_{i,t}^{\mathrm{pv},\psi})^2 \leq (\bar{s}_{i,t}^{\mathrm{pv},\psi})^2. \label{qpv}
	\end{gather}
\end{subequations}

%where $\mc{N}_{pv}$ denotes the set of buses connected with dispatchable PV units. %{\color{red} Here, we consider the voltage source inverter (VSI)-based PV units, which are dispatchable. }
% $P_{i,t}^{g,\psi}$ and $Q_{i,t}^{g,\psi}$ are the active and reactive PV generation power in phase $\psi$ of bus $i$ at time $t$, respectively. $\bar{P}_{i,t}^{g,\psi}$ and $\underline{P}_{i,t}^{g,\psi}$ are the upper and lower limits of active generation power. $\bar{S}_{i,t}^{g,\psi}$ denotes the apparent power capacity of PV units.

\subsubsection{Energy Storage Devices}
$\forall i \in \mc{N}_{\mathrm{es}}, \ t\in \mc{T}$
\begin{subequations} \label{bat}
   \begin{gather}
   \underline{p}_{i,t}^{\mathrm{es},\psi}  \leq  p_{i,t}^{\mathrm{es},\psi} \leq \bar{p}_{i,t}^{\mathrm{es},\psi} \label{es:powerlim}\\
   (p_{i,t}^{\mathrm{es},\psi})^2+(q_{i,t}^{\mathrm{es},\psi})^2 \leq (\bar{s}_{i,t}^{\mathrm{es},\psi})^2 \label{qpes}\\
   E^{\mathrm{es}}_{i,t}= \kappa_i \cdot E^{\mathrm{es}}_{i,t-1}-\Delta t\cdot p_{i,t}^{\mathrm{es}} \label{dyn}
   \\
    \underline{E}^{\mathrm{es}}_i  \leq E^{\mathrm{es}}_{i,t} \leq \bar{E}^{\mathrm{es}}_i, \   E^{\mathrm{es}}_{i,T}= E^{\mathrm{es}}_{i,0}. \label{soe}
\end{gather}
\end{subequations}
 Here, the active ES power output $p_{i,t}^{\mathrm{es},\psi}$ can be either positive (discharging) or negative (charging). In (\ref{dyn}),
 $\kappa_i\in(0,1]$ is the storage efficiency factor that models the energy loss over time, and $E^{\mathrm{es}}_{i,0}$ denotes the  initial state of charge (SOC). 
 We assume  $100\%$ charging and discharging energy conversion efficiency for simplicity, i.e., no power loss in the charging or discharging process. Constraint (\ref{soe}) imposes the SOC limits and requires that the final SOC $E^{\mathrm{es}}_{i,T}$ recovers the initial value for  sustainability.

\subsubsection{Directly Controllable Loads }
$\forall i \in \mc{N}_{\mathrm{dcl}}, \ t\in \mc{T}$
\begin{subequations} \label{conload} 
    \begin{gather}
       \underline{p}_{i,t}^{\mathrm{d},\psi} \leq  p_{i,t}^{\mathrm{d},\psi} \leq \bar{p}_{i,t}^{\mathrm{d},\psi},\ 
   q_{i,t}^{\mathrm{d},\psi}= \eta_{i}^{\mathrm{d}}\cdot p_{i,t}^{\mathrm{d},\psi}\label{conload:pow}  \\
   \underline{E}_i^{\mathrm{d}} \leq  \sum_{t\in \mathcal{T}} ( p_{i,t}^{\mathrm{d}}\cdot\Delta t) \leq \bar{E}_i^{\mathrm{d}}. \label{conload:ser} 
   \end{gather}
\end{subequations}
 In (\ref{conload:pow}), we assume fixed power factors with
constant $\eta_{i}^{\mathrm{d}}$. Equation (\ref{conload:ser})  ensures that the cumulative load energy consumption  lies within an accepted interval in order to complete the task \cite{optdr}, which imposes the quality of service constraint.   
%where $P_{i,t}^{d,\psi}$ and $Q_{i,t}^{d,\psi}$ are the active and reactive controlled loads in phase $\psi$ of bus $i$ at time $t$, respectively. $\bar{P}_{i,t}^{d,\psi}$ and $\underline{P}_{i,t}^{d,\psi}$ are the upper and lower limits for active load adjustment. Here, we assume constant $\eta_{i}^{d}$ that indicates fixed power factors.

\subsubsection{HVAC Systems}
$\forall i\in \mc{N}_{\mathrm{hv}}, \ t\in \mc{T}$
\begin{subequations} \label{hvac}
   \begin{gather}
      0 \leq  p_{i,t}^{\mathrm{hv},\psi} \leq \bar{p}_{i,t}^{\mathrm{hv},\psi},\
   q_{i,t}^{\mathrm{hv},\psi}=\eta_{i}^{\mathrm{hv}}\cdot p_{i,t}^{\mathrm{hv},\psi} \label{reac}\\
      \underline{F}_{i}\leq F^{\mathrm{in}}_{i,t} \leq \bar{F}_{i} \label{comt}\\
 F^{\mathrm{in}}_{i,t}=F^{\mathrm{in}}_{i,t-1}+\alpha_i\cdot\left(F^{\mathrm{out}}_{i,t}-F^{\mathrm{in}}_{i,t-1}\right)+{\Delta t}\cdot{\beta_i}  \cdot p_{i,t}^{\mathrm{hv}}.\label{temp}
\end{gather}
\end{subequations}
 In (\ref{reac}),  we use fixed power factors with
constant $\eta_{i}^{\mathrm{hv}}$.
 Equation (\ref{temp}) depicts the indoor temperature dynamics,
 where $\alpha_i\in(0,1)$ and $\beta_i$ are the parameters specifying the thermal characteristics of the buildings and the environment. A positive (negative)  $\beta_i$ indicates that the HVAC appliances work in the heating (cooling) mode, and $F^{\mathrm{in}}_{i,0}$ is the initial indoor temperature.
 See \cite{hvac} for  detailed explanations.

 \begin{remark} \normalfont
  To make a trade-off between  model precision and computational efficiency, we employ some commonly used approximation methods in building the DER models (\ref{pv})-(\ref{hvac}). Nevertheless, more realistic DER models  can be adopted if needed. For instance, a realistic ES model  that accounts for charging and discharging power loss can be formulated as 
  \begin{subequations} \label{eq:reales}
        \begin{align}
     & p_{i,t}^{\mathrm{es},\psi} = p_{i,t}^{\mathrm{dis},\psi} - p_{i,t}^{\mathrm{cha},\psi}, \ p_{i,t}^{\mathrm{dis},\psi}\geq 0,\ p_{i,t}^{\mathrm{cha},\psi}\geq 0\\
     & p_{i,t}^{\mathrm{dis},\psi}\cdot p_{i,t}^{\mathrm{cha},\psi} =0\label{eq:reales:comple}\\
     &  E^{\mathrm{es}}_{i,t}= \kappa_i \cdot E^{\mathrm{es}}_{i,t-1}-\Delta t\cdot \big(\frac{1}{\nu_i^{\mathrm{dis}}} \cdot p_{i,t}^{\mathrm{dis}} - \nu_i^{\mathrm{cha}}\cdot  p_{i,t}^{\mathrm{cha}} \big) \\
      & \text{Equations (\ref{es:powerlim}) (\ref{qpes}) (\ref{soe})}
  \end{align}
  \end{subequations}
  where $\nu_i^{\mathrm{cha}}\in (0,1]$ and $\nu_i^{\mathrm{dis}}\in (0,1]$  denote the charging and discharging efficiency coefficients, respectively. Equation (\ref{eq:reales:comple}) is the complementarity  constraint to ensure that the ES devices can not charge and discharge at the same time. The non-convexity issue caused by (\ref{eq:reales:comple}) can be addressed by the penalty reformulation approaches \cite{esexact}. For example, reference \cite{espen} removes the complementarity  constraint (\ref{eq:reales:comple}), and adds a penalty term 
  $\sum_{t\in\mathcal{T}}\sum_{i\in \mathcal{N}_{\mathrm{es}}} (p_{i,t}^{\mathrm{dis}}  + p_{i,t}^{\mathrm{cha}})\Delta t $ to the objective to  penalize simultaneous charging and discharging. In theory, as long as the DER constraints are convex, they can be included in the following power flexibility aggregation method, and the ARO framework and the CCG solution algorithm still work.

 \end{remark}

 \subsection{Power Flow and Network Model}

%Denote $\mathcal{N}_Y$ and $\mathcal{N}_\Delta$ as the set of buses with wye- and delta-connection respectively.Let column vector $\mb{s}_Y:=\mb{p}_Y+\jmath\,\mb{q}_Y\in \mathbb{C}^{3|\mathcal{N}_Y|}$ and $\mb{s}_\Delta:=\mb{p}_\Delta+\jmath\,\mb{q}_\Delta\in\mathbb{C}^{3|\mathcal{N}_\Delta|}$ collect all the three-phase complex power injection via wye- and delta-connection respectively. 

%Denote $N_Y$ and $N_\Delta$ as the number of buses via wye- and delta-connection respectively with $N_Y+N_\Delta=N$. 

For compact expression,  we stack all the three-phase controllable power injections at time $t$ into a long vector as
\begin{align}
\qquad \qquad\bm{x}_t:=   \left( p_{i,t}^{\mathcal{K},\psi},\; q_{i,t}^{\mathcal{K},\psi}    \right)_{i\in \mathcal{N},\, \mathcal{K}\in \{\mathrm{pv},\mathrm{es},\mathrm{d},\mathrm{hv}\},\,\psi  }.    \label{dex}
\end{align}

With the fixed-point linearization method introduced in \cite{linear_1},
we can derive the linear multi-phase power flow model (\ref{lp1})
based on a given operational point.
\begin{subequations} \label{lp1}
	\begin{align}
	\bm{v}_t \, &=\, \mb{A}\bm{x}_t\,\, +\bm{a}_t \label{v}\\
	\bm{i}_{t}\, &=\, \mb{B}\bm{x}_t\,\, +\bm{b}_t \label{il}\\
	p_{0,t}&=\, \bm{d}^\top\bm{x}_t+g_t \label{p_0}\\
		q_{0,t}&=\, \bm{f}^\top\bm{x}_t+h_t. \label{q_0}
	\end{align}
\end{subequations}
Here, matrices $\mb{A}, \mb{B}$, vectors $\bm{a}_t,\bm{b}_t,\bm{d},\bm{f}$ and scalars $g_t,h_t$ are all system parameters, whose detailed definitions and derivations are provided in  Appendix \ref{lpfm}. Note that $\bm{x}_t$ only contains the controllable DER power injection variables, while the time-varying  uncontrollable loads and non-dispatchable power generations are treated as given system parameters and captured by  $\bm{a}_t, \bm{b}_t, g_t,h_t$. 
In essence, the used power flow linearization method  can be viewed as a linear interpolation between two power flow solutions: the given operational point and a known operational point with no power injection. 
As a result, the  linear power flow model (\ref{lp1}) has better global approximation accuracy comparing with the standard linearized models based on local first-order Taylor expansion, and it is  applicable to both meshed and radial power networks.
In  \cite{linear_2}, a continuation analysis of the linear power flow model (\ref{lp1}) is performed on the IEEE 13-node system and a real feeder with about 2000 nodes, and the numerical results show that  the relative errors in voltages do not exceed 0.2\% and 0.6\%, respectively. 
Hence, the linear model (\ref{lp1}) provides an accurate approximation of unbalanced power flow for the proposed
method to achieve efficient  power flexibility aggregation.

%\textit{Remark 1:} For exposition  simplicity, we outline the linear power flow model (\ref{lp1}) for a three-phase system. However, the proposed framework is clearly applicable to the systems with a mix of three-, double-, single-phase connections. For example, if a electric device is  double-phase or single-phase integrated, we fix the entries of the missing phases as zero in $\{\mb{p}_Y,\mb{q}_Y\}$ or $\{\mb{p}_\Delta,\mb{q}_\Delta\}$  and the corresponding line impedance matrix. Besides, the linear power flow model (\ref{lp1}) captures all possible connection manners of electric devices, and it is applicable to both meshed and radial distribution networks.

Accordingly, the network constraints can be formulated as 
\begin{subequations}  \label{net_con}
	\begin{align} 
	\underline{\bm{v}} &\leq \bm{v}_t  \leq \bar{\bm{v}} \label{vol}\\
	\underline{\bm{i}} &\leq  \bm{i}_{t}\, \leq \bar{\bm{i}}  \label{thermal}
	\end{align}
\end{subequations}
which involve the voltage limit constraints  (\ref{vol}) and the line thermal constraints (\ref{thermal}).

\subsection{Comprehensive System Model} \label{sec:circularlinear}

% To facilitate the subsequent modelling with robust optimization, the linearization method introduced in \cite{rop3} is applied to approximate the circular apparent power capacity constraint (\ref{qpv}) (\ref{qpes}) with several concentric square constraints. As shown in Figure \ref{linearization}, the linear approximation achieves higher accuracy with more square constraints. Generally, two square constraints with the  rotation angle of 45$^\circ$ are  sufficiently accurate for practical applications. Hence, constraint (\ref{qpv}) of PV units can be approximated by 
%\begin{subequations}
%    \begin{gather}
%  -\bar{s}_{i,t}^{g,\psi}\leq   p_{i,t}^{g,\psi} \leq \bar{s}_{i,t}^{g,\psi} \\    -\bar{s}_{i,t}^{g,\psi}\leq   q_{i,t}^{g,\psi} \leq \bar{s}_{i,t}^{g,\psi} \\      -\sqrt{2}\,\bar{s}_{i,t}^{g,\psi}\leq   p_{i,t}^{g,\psi}+ q_{i,t}^{g,\psi}\leq \sqrt{2} \,\bar{s}_{i,t}^{g,\psi} \\     -\sqrt{2}\,\bar{s}_{i,t}^{g,\psi}\leq   p_{i,t}^{g,\psi}- q_{i,t}^{g,\psi}\leq \sqrt{2}\, \bar{s}_{i,t}^{g,\psi} 
%\end{gather}
%\end{subequations}
% which is similar for constraint (\ref{qpes}) of ES devices.

%\begin{figure}[thpb]
%	\centering
%	\includegraphics[scale=0.23]{circle.png}
%	\caption{Linearization method  for circular apparent power capacity constraint.}	\label{linearization}
%\end{figure}

Define $\bx := (\bx_t)_{t\in \mathcal{T}}$, $\bm{p}_0: = (p_{0,t})_{t\in \mathcal{T}}$, $\bm{q}_0: = (q_{0,t})_{t\in \mathcal{T}}$. Then the comprehensive system model, including the multi-period DER models  (\ref{pv})-(\ref{hvac}) and network model (\ref{lp1}) (\ref{net_con}), can be reformulated as the following compact   form (\ref{fin}):
\begin{subequations} \label{fin}
	\begin{gather} 
\bm{p}_{0} = \mb{D} \bx + \bm{g} \label{pp0}\\
\bm{q}_{0} = \mb{F} \bx + \bm{h} \label{qq0}\\
 ||\mb{E}_l \bx||\leq {s}_l, \ \forall l\in\mathcal{L} \label{eq:quadra}\\
\mb{W} \bx \leq \bm{w}. \label{els} 
	\end{gather}
\end{subequations}
Here, equation (\ref{pp0}) and (\ref{qq0}) are the stacks of equation (\ref{p_0}) and (\ref{q_0})  
for all time periods $t\in\mathcal{T}$, respectively. Equation (\ref{eq:quadra}) describes all the apparent power capacity constraints for PV units and ES devices, i.e., the square roots of both sides of (\ref{qpv}) (\ref{qpes}),
and $\mathcal{L}$ is the index set of these constraints.
Equation (\ref{els}) captures the remaining DER  and  network constraints, where  equalities are reformulated in an equivalent unified form as inequalities. Matrices $ \mb{D}, \mb{F}, \mb{E}_l, \mb{W}$, vectors $\bm{g}, \bm{h}, \bm{w}$ and scalar ${s}_l$ are the corresponding system parameters. Note that the SOC constraints  (\ref{dyn}) (\ref{soe}) and the temperature constraints (\ref{comt}) (\ref{temp}) are reformulated as  linear constraints on $\bx$ and contained in (\ref{els}). Taking  SOC constraints for example, we can eliminate the variable $E_{i,t}^{\mathrm{es}}$ and equivalently 
transform (\ref{dyn}) (\ref{soe}) as 
\begin{align*}
    \kappa_i^t\cdot  E_{i,0}^{\mathrm{es}} -\bar{E}^{\mathrm{es}}_i    \leq    \Delta t\cdot \sum_{\tau =1}^t (\kappa_i^{t-\tau}\cdot p_{i,\tau}^{\mathrm{es}}) \leq   \kappa_i^t\cdot  E_{i,0}^{\mathrm{es}} -\underline{E}^{\mathrm{es}}_i
\end{align*}
which is then included in the compact form (\ref{els}).

%% file: aggregation_method.tex
\section{Power Aggregation Methodology via Two-stage Adaptive Robust Optimization}

In this section, we interpret the power flexibility aggregation problem as the formulation in the ARO language, and propose two-stage ARO models to aggregate active  and reactive power flexibility for a distribution system. The practical application and power disaggregation are discussed as well.

\subsection{Power Aggregation Modelling via ARO} \label{sec:paggaro}

%Essentially, power flexibility aggregation can be regarded as the projection of high-dimensional network and DER operational constraints onto  the feasible region  of the aggregate power $(\bm{p}_0,\bm{q}_0)$. However, 

With massive DERs and multi-period power flow relation, the exact feasible region of the aggregate power $(\bm{p}_0,\bm{q}_0)$ is complex and intractable to procure or use. 
Instead, an  inner approximation is generally performed to obtain a concise and efficient representative of the exact feasible region. There are two desired properties of the approximate feasible region $\mathbb{D}$: %{\color{red}{change this $\mathbb{S}$ to be something else? It often used as sphere.}}

\begin{itemize}
    \item [1)] \emph{Aggregation optimality}: $\mathbb{D}$ is the optimal inner approximation of the exact feasible region with largest volume. 
    
    \item [2)] \emph{Disaggregation feasibility}: any aggregate power trajectory $(\bm{p}_0,\bm{q}_0)$ within  $\mathbb{D}$ can be fulfilled by the dispatch of DERs without violating operational constraints.
\end{itemize}

To achieve these two significant properties, the ARO framework can be leveraged to  formulate the power aggregation problem as model (\ref{eq:ARO}):
%As mentioned before, it is required to guarantee the property of disaggregation feasibility, so that any power regulation command determined by the upper-level power dispatch can be realized without violating the internal network and DER operational constraints. In other word, the power aggregation strategy should be robust to any scenario in the feasible region $\mathbb{S}$, while different scenarios are allowed to have different DER control schemes for fulfillment. This exactly matches the idea of ARO, thus the ARO framework is leveraged to formulate the power aggregation problem as model (\ref{eq:ARO}):
\begin{subequations} \label{eq:ARO}
    \begin{align} 
\text{Obj}.\ \ & \max_{\mathbb{D}} \ \text{volume}(\mathbb{D}) \label{eq:AROobj}\\
      \text{s.t.}\  \ &
   \forall  \left(\bp_0,\bq_0\right) \in \mathbb{D},\, \exists\, \bx(\bp_0,\bq_0 ) \label{subj}\\
   &\bp_0 = \mb{D} \bx(\bp_0,\bq_0 ) +\bm{g}\label{eq:ARO:p}\\
   &\bq_0 =\, \mb{F} \bx(\bp_0,\bq_0 ) +\bm{h} \label{eq:ARO:q}\\
& ||\mb{E}_l \bx(\bp_0,\bq_0 )||\leq {s}_l, \ \forall l\in\mathcal{L}  \label{eq:ARO:qua}\\
   &\mb{W} \bx(\bp_0,\bq_0 )\leq \bm{w} \label{eq:ARO:lin}
\end{align}
\end{subequations}
where objective (\ref{eq:AROobj}) aims to find the largest feasible region of the aggregate power to fully extract the DER flexibility. Equation (\ref{subj})  guarantees the disaggregation feasibility in an exact manner through the ARO modelling. It indicates that for any aggregate power trajectory $(\bp_0,\bq_0)$ within $\mathbb{D}$, there must exist a corresponding DER dispatch scheme $\bx$ to achieve it while respecting all the operational constraints.
In the ARO language, the (approximate) feasible region $\mathbb{D}$ is regarded as the uncertainty set, and $(\bp_0,\bq_0)$ is treated as the  uncertainty variable subject to the uncertainty set $\mathbb{D}$. The DER dispatch scheme  $\bx(\bp_0,\bq_0)$ refers to the adaptive variable that can be determined after the reveal of the uncertainty variable, thus it is  functional on $\bp_0$ and $\bq_0$. Instead of using a static variable $\bx$, the introduction of adaptive variables can significantly enhance the  optimality of the robust solutions \cite{aro}.

\begin{figure}
    \centering
    \includegraphics[scale= 0.38]{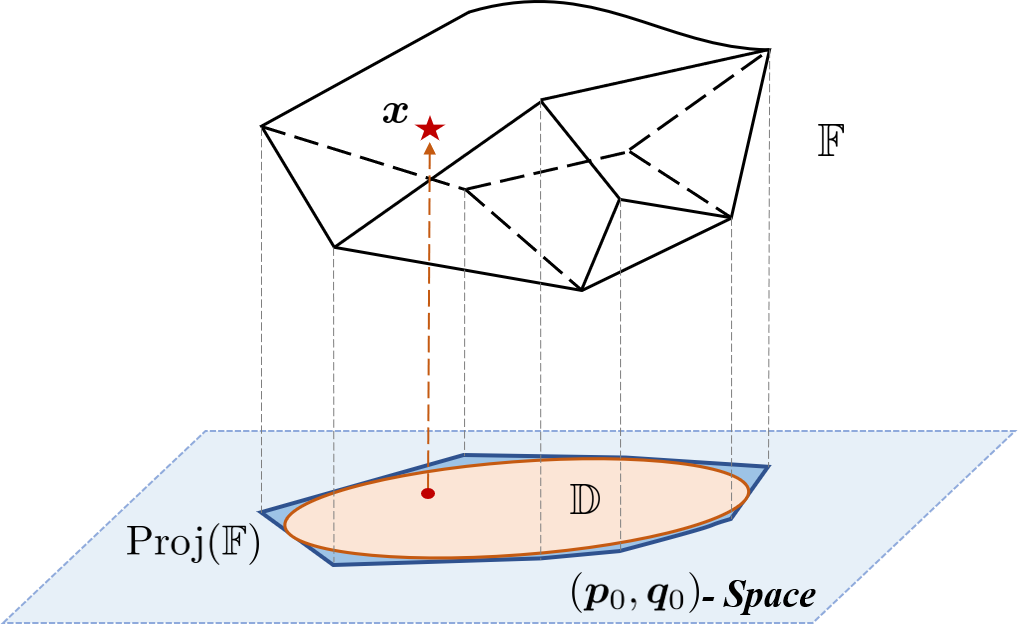}
    \caption{Illustration of leveraging ARO method for projection.}
    \label{fig:proj}
\end{figure}

\begin{remark}
\normalfont  
In essence, the power flexibility aggregation can be regarded as a projection of the high-dimensional feasible region of $\bx$ onto the low-dimensional space of the aggregate power $(\bm{p}_0,\bm{q}_0)$. As illustrated in Figure \ref{fig:proj}, 
the network and DER operational constraints (\ref{eq:ARO:qua}) (\ref{eq:ARO:lin}) constitute the high-dimensional feasible region $\mathbb{F}$ of $\bx$, and equations (\ref{eq:ARO:p}) (\ref{eq:ARO:q}) describe the mapping from $\bx$ to $(\bm{p}_0,\bm{q}_0)$. The ARO constraint (\ref{subj}) enforces that the feasible set $\mathbb{D}$ on  $(\bm{p}_0,\bm{q}_0)$-space is inscribed within the exact projected set, i.e., $\mathbb{D}\subseteq\mathrm{Proj}(\mathbb{F})$, since for any point $(\bm{p}_0,\bm{q}_0)\!\in\!\mathbb{D}$, there exists a corresponding $\bx(\bm{p}_0,\bm{q}_0)\!\in\! \mathbb{F}$ such that the mapping (\ref{eq:ARO:p}) (\ref{eq:ARO:q}) is fulfilled. This condition is depicted in model (\ref{eq:ARO}) by means of the adaptive variables. Objective (\ref{eq:AROobj}) is defined to find the optimal feasible set $\mathbb{D}^*$ with  maximum volume. In this way, the proposed  model (\ref{eq:ARO}) is a general framework that can be extended to solve other high-dimensional projection problems.
\end{remark} 

%In practical application, by solving model (\ref{eq:ARO}),  a distribution system can obtain its optimal feasible region $\mathbb{D}^*$, and then participates in transmission-level power dispatch or provides reserve services  with $\mathbb{D}^*$. The disaggregation feasibility ensures that for any  power tracking commands $(\tilde{\bp}_0,\tilde{\bq}_0)\!\in\! \mathbb{D}^*$ issued by the upper-level dispatch, the distribution system is able to achieve the net power injection  $(\tilde{\bp}_0,\tilde{\bq}_0)$ by appropriately controlling the DER devices, i.e., determining the associated $\bx$.  The determination of optimal $\bx$ to execute the power tracking commands is referred as \emph{power disaggregation}, which is elaborated in Section III-D.

Model (\ref{eq:ARO}) is a general formulation of the power flexibility aggregation problem. 
In the following text, two concrete aggregation models with two-stage optimization are developed: one computes the time-decoupling optimal feasible intervals  for the aggregate active (or reactive) power, while the other solves the optimal elliptical feasible regions for the aggregate P-Q domain. Basically, the  feasible region $\mathbb{D}$ can be chosen as any  convex set, including time-coupling ellipsoid or polyhedron,  as long as it can be properly parameterized.

%Note that although the feasible region $\mathbb{S}$ is defined in a time-decoupled form, the determination of the optimal $\mathbb{S}$ need to coordinate all time periods  due to the existence of time-coupling DER devices, such as energy storage and HVAC systems. 
%Hence, we build two  optimization models below to perform optimal power aggregation: one aims to achieve the maximal flexibility level of the network system, while the other is established to optimally schedule the power flexibility reserve and the DER power dispatch for the base operational point.

\subsection{Active Power Aggregation Model }

This subsection focuses on characterizing the feasible region of the  exchanged active power at the substation interface, which is also applicable to aggregate reactive power.
To avoid computational burden and facilitate high-level application, we use the time-decoupling feasible intervals $\mathbb{D}_p$ to depict the feasible region of the aggregate active power %{\color{red}{change this $\mathbb{S}$ to be something else? It often used as sphere.}}: 
\begin{align} \label{inter}
\begin{split}
\mathbb{D}_p:=\left[{p}^\vee_{0,1}, {p}^\wedge_{0,1}\right]\times\left[{p}^\vee_{0,2}, {p}^\wedge_{0,2}\right]\times\dots\times\left[{p}^\vee_{0,T}, {p}^\wedge_{0,T}\right] 
\end{split}
\end{align}
where the notations with the superscript ``$\wedge$" and ``$\vee$" denote the upper and lower power bounds respectively. Accordingly, the possible exchanged active power at substation over time is restrained by the upper power trajectory $\bp_0^\wedge: = (p_{0,t}^\wedge)_{t\in \mathcal{T}}$ and the lower  power trajectory $\bp_0^\vee: = (p_{0,t}^\vee)_{t\in \mathcal{T}}$.

 Since the  feasible region $\mathbb{D}_p$ is described with the decision variables $(\bp_0^\vee,\bp_0^\wedge)$, 
a fixed uncertainty set is desired to replace it in order to fit the ARO  framework. To this end, we introduce $\bxi\!:=\!(\xi_t)_{t\in\mathcal{T}}\!\in\! \R^{T}$  as the uncertainty variable  and formulate the actual aggregate active power as 
\begin{align} \label{eq:p0xi}
    p_{0,t} = \xi_t\cdot p_{0,t}^\wedge +(1-\xi_t)\cdot p_{0,t}^\vee, \quad \forall t\in\mathcal{T}
\end{align}
which can be regarded as a linear combination between the lower bound $p_{0,t}^\vee$ and the upper bound $p_{0,t}^\wedge$ with weight $\xi_t\in[0,1]$.  In particular, $p_{0,t}$ equals to $p_{0,t}^\wedge$ when $\xi_t=1$, and  equals to  $p_{0,t}^\vee$ when $\xi_t=0$.  In this way, by the change of variables, we can equivalently replace the robust condition ``$\forall \bm{p}_0\!\in\! \mathbb{D}_p$" by ``$\forall \bxi\!\in \!\mathbb{U}_1$", where $\mathbb{U}_1$ is the fixed box uncertainty set (\ref{eq:U1}):
\begin{align}\label{eq:U1}
    \mathbb{U}_1: = \{\bxi\,|\, 0\leq \xi_t\leq 1,\ \forall t\in \mathcal{T}\}.
\end{align}

As a result, based on the general ARO formulation (\ref{eq:ARO}), we develop the   active power aggregation (\textbf{APA}) model  as (\ref{eq:APA})
to obtain the optimal $\mathbb{D}_p$ with maximal flexibility.
\begin{subequations} \label{eq:APA}
    \begin{align}
%   & (\textbf{APA Model}) \nonumber \\
   \text{Obj}. \ \ &  \max_{\bp^\wedge_0,\bp^\vee_0} \, \min_{\bxi\in\mathbb{U}_1} \, \max_{\bx(\bxi)}\  \bm{1}^\top \left( \bp^\wedge_0 -\bp^\vee_0   \right) \label{eq:APA:obj}\\
        \text{s.t.} \  \  & \bp^\vee_0\leq \bp^\wedge_0\label{eq:APA:p0}\\
         &   \bm{\xi} \circ \bp^\wedge_0 +(\bm{1}-\bxi)\circ\bp^\vee_0   = \mb{D}\bx(\bm{\xi}) + \bm{g},  \,   \forall \bxi \in \mathbb{U}_1  \label{eq:APA:px} \\
        &  ||\mb{E}_l \bx(\bm{\xi} )||\leq {s}_l, \qquad\qquad\qquad\,   \forall l\in\mathcal{L},\,\bxi \in \mathbb{U}_1\\
        &\mb{W}\bx(\bm{\xi}) \leq \bm{w},\qquad\qquad\qquad\qquad\quad \ \ \; \forall \bxi \in \mathbb{U}_1 \label{eq:APA:netcon}
      %\\  & \mathbb{U}_p : = [0,1]\times [0,1]\times \cdots \times [0,1] \label{eq:APA:uncerset}
    \end{align}
\end{subequations}
where the left-hand side of (\ref{eq:APA:px}) is  the vector version of (\ref{eq:p0xi}).
The objective (\ref{eq:APA:obj}) is in the form of two-stage optimization. In the first stage,  $(\bp_0^\wedge,\bp_0^\vee)$ is the ``here-and-now"  decision that maximizes the total aggregate flexibility. In the second stage, the 
DER power dispatch scheme $\bx(\bxi)$ denotes the ``wait-and-see" adaptive decision that can be made after the uncertainty variable $\bxi$ is revealed. Through the minimax optimization, it ensures that for the worst-case scenarios in $\mathbb{U}_1$, 
there exists a corresponding feasible $\bx(\bxi)$ to fulfill them. In this manner, the disaggregation feasibility of the solved feasible intervals is guaranteed with optimality.

Besides, the APA model (\ref{eq:APA}) can be modified with different objectives and settings to meet the power aggregation goals. For example, by adding a base power trajectory, it can formulate an economic dispatch model that optimally schedules the flexibility reserve and DER power for the distribution system. See the economic power aggregation model in \cite{our1} for details.

\begin{remark} \normalfont
In our previous work \cite{our1}, two 
  DER power dispatch schemes $\{\bx^\vee,\bx^\wedge\}$ associated with the lower and upper aggregate power trajectories $\{\bp_0^\vee, \bp_0^\wedge\}$ are considered in the flexibility aggregation model. And the
 heuristic constraints (\ref{eq:heu}) on the ES outputs and HVAC loads are imposed to ensure  disaggregation feasibility. See \cite[Appendix B]{our1} for details. 
\begin{subequations} \label{eq:heu}
    \begin{align}
 \qquad\qquad  & p_{i,t}^{\mathrm{es},\wedge}\leq p_{i,t}^{\mathrm{es},\vee}, & \forall i\in \mathcal{N}_{\mathrm{es}}, t\in\mathcal{T}\  \\
 \qquad \qquad & p_{i,t}^{\mathrm{hv},\vee}\leq p_{i,t}^{\mathrm{hv},\wedge}, &\forall i\in \mathcal{N}_{\mathrm{hv}}, t\in\mathcal{T}.
\end{align}
\end{subequations}
However, constraints (\ref{eq:heu}) are sufficient but unnecessary conditions to achieve disaggregation feasibility, thus may lead to sub-optimal solutions. In contrast,
the APA model (\ref{eq:APA}) guarantees  disaggregation feasibility  through the  adaptive robust constraints (\ref{eq:APA:px})-(\ref{eq:APA:netcon}) in an exact way, instead of using  conservative  heuristic constraints (\ref{eq:heu}),
therefore the  optimality of solutions is enhanced and the power flexibility can be fully exploited. This is also validated by  the numerical comparisons carried out in  Section V-A.
%The numerical comparisons  carried out in  Section \ref{sec:sim:com} also validate the superiority of the proposed  ARO method over    work \cite{our1}. 
In addition, it is not clear how to generalize the heuristic constraints (\ref{eq:heu}) used in \cite{our1} to aggregate both active and reactive power, while  the proposed ARO method  is clearly applicable to the P-Q aggregation problem, which is elaborated in the next subsection.

\end{remark}

\subsection{Active-Reactive Power Aggregation Model}

In power systems, reactive power plays a pivotal role in maintaining voltage security and reducing network loss.
The flexibility of reactive power from massive DERs can also be exploited to support the bulk transmission system.
Since active and reactive power are highly coupled in both the network constraints and DER operational constraints, it necessitates a joint P-Q flexibility aggregation scheme. 
As investigated in \cite{opf1,ellip}, the elliptical feasible region\footnote{Essentially, except elliptical feasible regions, other parameterized convex sets, such as polygons, can also be used to depict the aggregate P-Q flexibility, and the proposed power aggregation method is applicable as well.} is a good choice to depict the snapshot of the aggregate P-Q flexibility at each time. 
Hence, we parameterize the feasible region $\mathbb{D}_{pq}$ with time-decoupling ellipses:
\begin{align}
    \mathbb{D}_{pq}: = \prod_{t \in\mathcal{T}} \left\{ 
    \begin{bmatrix}p_{0,t}\\q_{0,t}
    \end{bmatrix} =  \begin{bmatrix}p_{0,t}^c\\q_{0,t}^c
    \end{bmatrix} +  \bm{Y}_t\cdot\bxi_t: \,  ||\bxi_t||\leq 1 \right\}
\end{align}
where the parameter $(p_{0,t}^c, q_{0,t}^c)$ denotes the center and the $2\times 2$-dimension positive semidefinite matrix $\bm{Y}_t$ describes the rotation and stretch transformation for the ellipse \cite{max}.

Accordingly, the active-reactive power aggregation (\textbf{ARPA}) model (\ref{eq:ARPA}) is built to optimally aggregate the P-Q flexibility:
\begin{subequations} \label{eq:ARPA}
    \begin{align}
        \text{Obj.} \ &\max_{p^c_{0,t},q^c_{0,t},\bm{Y}_t} \, \min_{\bxi\in\mathbb{U}_2} \, \max_{\bx(\bxi)}\   \sum_{t\in \mathcal{T}} \log(\text{det}(\bm{Y}_t)) \label{eq:ARPA:obj}\\
        \text{s.t.} \  &\  \bm{Y}_t\succeq 0,\ \bxi_t\in\mathbb{R}^2\qquad\qquad\ \ \forall t\in\mathcal{T} \label{eq:ARPA:semi}\\
        \begin{split}
                 &          \begin{bmatrix}
        p_{0,t}^c \\ 
         q_{0,t}^c 
       \end{bmatrix} + \bm{Y}_t\cdot \bxi_t = 
       \begin{bmatrix}
       \bm{d}^\top \\ \bm{f}^\top
       \end{bmatrix}\bx_t(\bxi) +
       \begin{bmatrix}
       g_t\\h_t
       \end{bmatrix} \\
  &     \qquad\qquad\qquad\qquad\qquad\qquad \forall t\in \mathcal{T},\bxi \in \mathbb{U}_2
        \end{split}\label{eq:ARPA:pq}\\
         & ||\mb{E}_l \bx(\bm{\xi} )|| \leq {s}_l, \qquad\qquad\quad\ \forall l\in\mathcal{L},\,\bxi \in \mathbb{U}_2\\
        &\ \mb{W}\bx(\bm{\xi}) \leq \bm{w},\qquad\qquad\quad \ \ \, \forall \bxi \in \mathbb{U}_2. \label{eq:ARPA:netcon}
    \end{align}
\end{subequations}
Here, $\bxi := (\bxi_t)_{t\in\mathcal{T}}$ denotes the uncertainty variable subject to the fixed uncertainty set $\mathbb{U}_2$ (\ref{eq:ARPA:uncerset}):
\begin{align} \label{eq:ARPA:uncerset}
      \mathbb{U}_2 : = \left\{\bxi\, | \, ||\bxi_t||\leq 1,\ \forall t\in \mathcal{T}\right\}. 
\end{align} 
Similarly, we can equivalently replace the robust condition ``$\forall(\bp_0,\bq_0)\in \mathbb{D}_{pq}$" by ``$\forall \bxi\in \mathbb{U}_2$"  with equation (\ref{eq:ARPA:pq}).

The time-variant elliptical feasible regions with parameters $(p_{0,t}^c,q_{0,t}^c,\bm{Y}_t)_{t\in\mathcal{T}}$ are the ``here-and-now" decisions, while the DER power dispatch scheme $\bx(\bxi)$ denotes the  ``wait-and-see" adaptive variable that can be determined after the uncertainty variable $\bxi$ is revealed. 
In objective (\ref{eq:ARPA:obj}), $\text{det}(\bm{Y}_t)$ denotes the determinant of matrix $\bm{Y}_t$, which equals to 
the area of the ellipse \cite{max} at time $t$.
Thus the objective (\ref{eq:ARPA:obj}) aims to maximize the total aggregate flexibility of active and reactive power over $T$ time periods.
Using the ARO constraints (\ref{eq:ARPA:pq})-(\ref{eq:ARPA:netcon}), the disaggregation feasibility is guaranteed with exactness. With the semi-definite constraint in (\ref{eq:ARPA:semi}), the ARPA model (\ref{eq:ARPA}) is  a two-stage   semi-definite programming (SDP) problem.

%\begin{remark}
%To facilitate practical application, we use time-decoupling intervals and ellipsoids to describe the feasible region of active and reactive power in  APA model (\ref{eq:APA}) and ARPA model (\ref{eq:ARPA}) respectively, so that the  aggregate feasibility  is defined in a simple and concise form with few critical parameters, which is efficient for upper-level power dispatch decision-making. Nevertheless, more complicated and detailed feasible regions can be defined and used to describe the aggregate power flexibility if necessary,  while the ARO framework (\ref{eq:ARO}) is still applicable to such cases.
%\end{remark}

As indicated in objectives (\ref{eq:APA:obj}) and (\ref{eq:ARPA:obj}), the main difference between the APA model (\ref{eq:APA}) and the ARPA model  (\ref{eq:ARPA}) is that APA aims to compute the optimal feasible region $\mathbb{D}_p^*$ of the aggregate active power $\bp_0$, while ARPA aims to procure the optimal aggregate feasible region $\mathbb{D}_{pq}^*$ of both  $\bp_0$ and $\bq_0$. Note that the projection of $\mathbb{D}_{pq}^*$  onto the $\bp_0$-space is not necessary to be $\mathbb{D}_p^*$. Actually, these two power flexibility aggregation models are proposed for distinct application scenarios. In most of present electricity markets, only the active power is traded and dispatched by the independent system operator (ISO) in the transmission level. While the reactive power in a distribution system  is generally controlled locally   for voltage regulation and network loss mitigation. Accordingly, 
the APA model (\ref{eq:APA}) is designed to aggregate only the active power and participate in the transmission-level power dispatch with the feasible region $\mathbb{D}_p^*$.
However, as the penetration of inverter-based DERs that can independently control the outputs of active power and reactive power \cite{incor} increases rapidly, it is realized that the reactive power flexibility of a distribution system can also be aggregated to support the transmission system operation \cite{td-3}, e.g., 
providing ancillary services. To this end, the ARPA model (\ref{eq:ARPA}) is designed to aggregate both active power and reactive power.

\subsection{Practical Application and Power Disaggregation}

%In practical application, by solving model (\ref{eq:ARO}),  a distribution system can obtain its optimal feasible region $\mathbb{D}^*$, and then participates in transmission-level power dispatch or provides reserve services  with $\mathbb{D}^*$. The disaggregation feasibility ensures that for any  power tracking commands $(\tilde{\bp}_0,\tilde{\bq}_0)\!\in\! \mathbb{D}^*$ issued by the upper-level dispatch, the distribution system is able to achieve the net power injection  $(\tilde{\bp}_0,\tilde{\bq}_0)$ by appropriately controlling the DER devices, i.e., determining the associated $\bx$.  The determination of optimal $\bx$ to execute the power tracking commands is referred as \emph{power disaggregation}, which is elaborated in Section III-D.

In practice, by solving the APA (or ARPA) model, a distribution system can obtain the optimal aggregate feasible region, and then  participates in transmission-level power scheduling and dispatch as a virtual power plant. In this process, the feasible region of a distribution system is used in the same way as the capability curves of conventional generators. Moreover, since the DER devices are mostly power electronics  inverter-interfaced, the aggregate power can be adjusted  fast with a high ramping rate
in response to the regulation signals. Therefore, the distribution system with the aggregate feasible region is capable of providing ancillary services, including regulation service and reserve service, to the bulk transmission system. We take the frequency regulation service as an example\footnote{With an additional base power trajectory, providing reserve service with the aggregate feasible region is similar to the procedure of providing regulation service, which is introduced in  our previous work \cite[Section  IV-A]{our1} in details.} and describe the specific implementation process as follows. 
\begin{itemize}
	\item [1)] Each distribution system performs the APA model (\ref{eq:APA}), and reports its optimal feasible intervals $\mathbb{D}_p^*\!:=\!\left[p_{0,t}^{\vee*}, p_{0,t}^{\wedge*} \right]_{t\in\mathcal{T}} $ to the transmission system operator.
	\item[2)] Based on the reported aggregate feasible intervals of all  distribution systems and the capability curves of all  generators, the 
	transmission system operator conducts a holistic power dispatch and determines the associated  regulation power trajectory $\bp_0^{\mathrm{reg}}\!:=\!(p_{0,t}^{\mathrm{reg}})_{t\in\mathcal{T}}\!\in\! \mathbb{D}_p^*$ for each distribution system. 
	\item [3)]  Once receiving the regulation power trajectory $\bp_0^{\mathrm{reg}}$, each distribution system solves the following \emph{power disaggregation} problem (\ref{eq:PD}) to determine the optimal DER dispatch scheme $\bx^*$, then executes this scheme to track the regulation trajectory. 
\end{itemize}
The power disaggregation (PD) model (\ref{eq:PD}) is formulated as 
\begin{subequations} \label{eq:PD}
		\begin{align} 
		\text{Obj.}& \ \ \min_{\bm{x}} \ \sum_{t\in\mathcal{T}}  C_t\left( \bm{x}_t \right) %+ \rho_t\cdot(p_{0,t}-p_{0,t}^{\mathrm{reg}} )^2    \right]
		  \label{pd:obj_e}\\
		\text{s.t.}&\  \ \bm{p}_{0}^{\mathrm{reg}} = \mb{D} \bx + \bm{g},\  \text{ Equation  (\ref{eq:quadra}) (\ref{els})} \label{eq:pd:con}
		\end{align}
	\end{subequations}
where 	$C_t(\cdot)$ is the operational cost function of the distribution system, and a paradigmatic formulation is given by (\ref{eq:cost}) \cite{our1}.
The PD  model (\ref{eq:PD}) aims to minimize the total cost under the network and DER operational constraints while tracking the regulation power trajectory $\bp_0^{\mathrm{reg}}$.
	 Due to the 
	 disaggregation feasibility of the proposed power aggregation method, the PD model (\ref{eq:PD}) must be feasible for any $\bp_0^{\mathrm{reg}}\!\in\! \mathbb{D}_p^*$.
\begin{align}
\begin{split}
C_t&(\bx_t):=  \sum_{i\in\mathcal{N}_{\mathrm{es}}}\! c_{i}^{\mathrm{es}}\cdot(p_{i,t}^{\mathrm{es}})^2 +\!\sum_{i\in\mathcal{N}_{\mathrm{hv}}} \! c_i^{\mathrm{hv}}\cdot(F_{i,t}^{\mathrm{in}} - F_i^{\mathrm{cf}})^2 \\
&+\!\sum_{i\in\mathcal{N}_{\mathrm{pv}}}\left[ c_{1,i}^{\mathrm{pv}}\cdot{p_{i,t}^{\mathrm{pv}}}+ c_{2,i}^{\mathrm{pv}}\cdot({p_{i,t}^{\mathrm{pv}}-\bar{p}_{i,t}^{\mathrm{pv}}})^2  \right] +c_t\cdot p_{0,t}.  \label{eq:cost}
\end{split}
\end{align}
In (\ref{eq:cost}), the first term captures the damaging effect of  charging/discharging to the ES devices. The second term describes the HVAC disutility of deviating from the most comfortable temperature $F_i^{\mathrm{cf}}$. The third term denotes the operational cost and the power curtailment cost of PV units. $c_{i}^{\mathrm{es}}, c_{i}^{\mathrm{hv}}, c_{1,i}^{\mathrm{pv}},c_{2,i}^{\mathrm{pv}}$ are the corresponding cost coefficients. The last term is the cost of purchasing electricity from the transmission grid with the real-time price $c_t$.

%% file: Solution_alg.tex
\section{Solution Algorithm}

Since both the APA model (\ref{eq:APA}) and ARPA model (\ref{eq:ARPA}) are two-stage ARO problems, 
this paper employs the 
widely-used column-and-constraint generation (CCG) algorithm \cite{ccg} to solve them. According to the different decision-makings in the two stages, the original ARO model is decomposed as a master problem and a sub-problem, then a master-sub iterative process is performed to obtain the optimal solution. Taking the APA model (\ref{eq:APA}) for example, the CCG solution algorithm is presented as follows, while the solution method for the ARPA model (\ref{eq:ARPA}) is similar.

\subsection{Master Problem}
Following the decomposition structure of CCG algorithm, the master problem  is developed as (\ref{eq:master}), which corresponds to the first-stage decision making in the APA model (\ref{eq:APA}).
\begin{subequations} \label{eq:master}
    \begin{align}
    %  & (\textbf{Master-APA}) \nonumber \\
        \text{Obj.} \ \ &f_M = \max_{\bp_0^\wedge,\bp_0^\vee,\bx^k} \   \bm{1}^\top \left( \bp^\wedge_0 -\bp^\vee_0   \right)\\
        \text{s.t.} \  \  & \text{Equation (\ref{eq:APA:p0})}\\
      \begin{split}
 &      \bm{\xi}_*^k \circ \bp^\wedge_0 +(\bm{1}-\bxi_*^k)\circ\bp^\vee_0 = \mb{D}\bx^k + \bm{g}, \\
      & \qquad\qquad\qquad\qquad\qquad\,   \forall k =1,2,\cdots,K 
     \end{split} \\
        & ||\mb{E}_l \bx^k||\leq {s}_l, \quad \ \,
    \forall l\in\mathcal{L},\,  k =1,2,\cdots,K \\
        & \mb{W}\bx^k \leq \bm{w}, \qquad\qquad\quad\,  \forall k =1,2,\cdots,K
    \end{align}
\end{subequations}
where $(\bxi_*^k)_{k=1,2,\cdots,K}$ are given as known parameters.

Essentially, the master problem (\ref{eq:master}) can be regarded as a multi-scenario relaxation of the original two-stage APA model (\ref{eq:APA}). In particular, the 
uncertainty set $\mathbb{U}_1$ in (\ref{eq:APA}) is replaced by $K$ enumerated scenarios $(\bxi_*^k)_{k=1,2,\cdots,K}$ within $\mathbb{U}_1$, and each scenario is assigned a  corresponding  $\bx^k$ for adaptivity. Since finite enumerations are used  in (\ref{eq:master}) instead of the entire uncertainty set, the objective value $f_M$ offers an upper bound for the original APA model (\ref{eq:APA}).
As more and more scenarios and constraints are added, $f_M$ is expected to decrease to the optimal objective value of the APA model (\ref{eq:APA}). 

%The master problem (\ref{eq:APA}) of APA model is a quadratic constrained programming. Using the same decomposition method, the master problem of ARPA model (\ref{eq:ARPA}) becomes a quadratic constrained SDP, whose detailed formulation is omitted here. 
%Both of the master problems are convex optimization and thus can be solved efficiently.

\subsection{Sub-Problem}

The sub-problem (\ref{eq:sub}) is associated with the second-stage decision making in the APA model (\ref{eq:APA}), which optimizes $\bxi$ and $\bx(\bxi)$ with given $(\bp^{\vee}_{0*},\bp^{\wedge}_{0*})$:
\begin{subequations} \label{eq:sub}
    \begin{align}
    %  & (\textbf{Sub-APA}) \nonumber \\
        \text{Obj.} \ \ &f_S = \min_{\bxi\in \mathbb{U}_1} \max_{\bx(\bxi)}\ 0 \\
        \text{s.t.} \  \  
              &  \bm{\xi} \circ \bp^\wedge_{0*}+ (\bm{1} - \bxi)\circ \bp^\vee_{0*}  = \mb{D}\bx(\bxi) + \bm{g}\label{eq:sub:p0}\\
        &||\mb{E}_l \bx(\bxi)||\leq {s}_l, \quad \forall l\in\mathcal{L}  \label{eq:sub:qua}\\
        & \mb{W}\bx(\bxi) \leq \bm{w}. \label{eq:sub:netcon}
    \end{align}
\end{subequations}

With  given $(\bp^{\vee}_{0*},\bp^{\wedge}_{0*})$, if there exists a certain extreme scenario $\bxi\in \mathbb{U}_1$ such that no 
 corresponding feasible $\bx$ satisfies constraints (\ref{eq:sub:p0})-(\ref{eq:sub:netcon}), then the optimal objective value $f_S$ is $-\infty$;  otherwise $f_S =0$. Hence,  the sub-problem (\ref{eq:sub}) serves as a judge to determine whether the optimal feasible interval $[\bp^{\vee}_{0*},\bp^{\wedge}_{0*}]$ generated by  the master problem (\ref{eq:master}) guarantees the disaggregation feasibility. 
In addition, its optimal solution $\bxi_*$ is regarded as  the worst-case scenario  that jeopardizes the disaggregation feasibility, and thus can be 
 added to the master problem (\ref{eq:master}) as an enumerated scenario to improve the master solution. 
% The sub-problem of the ARPA model (\ref{eq:ARPA}) is similar to (\ref{eq:sub}), with constraint (\ref{eq:sub:p0}) replaced by the associated constraint of (\ref{eq:ARPA:pq}).

\subsection{Solution Method for Sub-Problem}

 The sub-problem (\ref{eq:sub}) is a minimax bi-level optimization. To solve it, the following reformulation technique is used to obtain a tractable optimization model.
Firstly, through strong duality on the inner maximization, the sub-problem (\ref{eq:sub})  can be reformulated as the monolithic optimization form (\ref{eq:dual}): 
\begin{subequations} \label{eq:dual}
    \begin{align}
    \begin{split} \label{eq:dual:obj}
         \text{Obj.}  \ &\min_{\bxi,\, \bm{\mu},\,  \bm{\lambda},\bm{\gamma}_l, \sigma_l}\  (\bp^\wedge_{0*}- \bp^\vee_{0*})^\top (\bm{\mu}\circ \bxi ) + \bm{w}^\top \bm{\lambda} \\
         &\qquad\qquad\qquad + (\bp^{\vee}_{0*}-\bm{g})^\top \bm{\mu}+ \sum_{l\in\mathcal{L}}s_l\sigma_l   
    \end{split}
       \\
     \text{s.t.}  \ & \quad
                \mb{D}^\top \bmu  + \sum_{l\in\mathcal{L}} \mb{E}_l^\top \bga_l + \mb{W}^\top\bm{\lambda} = \bm{0} \label{eq:dual:con} \\
                & \quad ||\bm{\gamma}_l||\leq \sigma_l,\quad \forall l\in\mathcal{L}\label{eq:socp}\\
                &\quad  \bxi\in \mathbb{U}_1, \ \bm{\lambda} \geq \bm{0}
    \end{align}
\end{subequations}
where  $\bm{\mu},\bm{\lambda},(\bm{\gamma}_l,\sigma_l)_{l\in\mathcal{L}}$ are all dual variables. The detailed derivation of (\ref{eq:dual}) is provided in Appendix \ref{app:duality}.

In the objective (\ref{eq:dual:obj}), there exists a nonconvex bilinear term $\bm{\mu}\circ \bxi$ that complicates the solution. Fortunately, the feasible regions of $\bmu$ and $\bxi$ are disjoint, which leads to the fact that the optimality of model (\ref{eq:dual}) must be achieved at one extreme point of the box uncertainty set $\mathbb{U}_1$ \cite{rop1}. It means that  the optimal $\xi_{t}$ must equal to $0$ or $1$ for all $t$, thus we can force $\xi_t$ to be a binary variable $\xi_t\in\{0,1\}$ without loss of optimality.
Then, using the big-M method,  the bilinear term $\bm{\mu}\circ \bxi$ can be linearized with the following two steps:

1) Define new non-negative variables $\bm{\mu}^+\!:=\!(\mu_t^+)_{t\in\mathcal{T}}\geq 0,\bm{\mu}^-\!:=\!(\mu_t^-)_{t\in\mathcal{T}}\geq 0$ to substitute $\bmu$ with $\bmu = \bmu^+ - \bmu^-$.

2)  Introduce new variables  $\bm{\nu}^+\!:=\!(\nu_t^+)_{t\in\mathcal{T}},\, \bm{\nu}^-\!:=\!(\nu_t^-)_{t\in\mathcal{T}}$  to substitute  the resultant products $\bm{\mu}^+\circ \bxi,\, \bm{\mu}^-\circ \bxi$, respectively. And constraint (\ref{eq:relax}) is  added to make this substitution equivalent. 
\begin{subequations} \label{eq:relax}
        \begin{align}
           & \bm{0}\leq \bm{\nu}^+\leq \bm{\mu}^+, \ \bm{\mu}^+\!- M(\bm{1}-\bxi) \leq \bm{\nu}^+\leq M\bxi \label{eq:tran:con+}\\
             & \bm{0}\leq \bm{\nu}^-\leq \bm{\mu}^-, \ \bm{\mu}^- \!- M(\bm{1}-\bxi) \leq \bm{\nu}^-\leq M\bxi \label{eq:tran:con-}
        \end{align}
\end{subequations}
where $M$ is a large positive value. When $\xi_t = 1$, constraint (\ref{eq:relax}) leads to $\nu^+_t = \mu_t^+, \nu^-_t = \mu_t^-$. When $\xi_t = 0$, it leads to $\nu^+_t = 0, \nu^-_t = 0$.

 As a consequence, the dual sub-problem (\ref{eq:dual}) is equivalently reformulated as model (\ref{eq:trandual}):
 \begin{subequations} \label{eq:trandual}
    \begin{align}
    \begin{split}
         \text{Obj.}  \ &\min_{\bxi,\, \bm{\mu},\, \bm{\nu}, \bm{\lambda},\bm{\gamma}_l, \sigma_l}\  (\bp^\wedge_{0*}- \bp^\vee_{0*})^\top (\bm{\nu}^+-\bm{\nu}^- )  + \bm{w}^\top \bm{\lambda} \\
         &\qquad + (\bp^{\vee}_{0*}-\bm{g})^\top (\bm{\mu}^+-\bm{\mu}^-)+ \sum_{l\in\mathcal{L}}s_l\sigma_l   
    \end{split}
       \\
     \text{s.t.}  \ &\quad \text{Equation (\ref{eq:socp}) (\ref{eq:relax})} \\ 
   &  \quad \mb{D}^\top (\bmu^+-\bmu^-)  + \sum_{l\in\mathcal{L}} \mb{E}_l^\top \bga_l + \mb{W}^\top\bm{\lambda} = \bm{0}  \\
                &\quad  \bxi\in \{0,1\}^T,\, \bm{\lambda}\geq \bm{0},\,\bmu^+\geq \bm{0},\,\bmu^- \geq \bm{0}.
    \end{align}
\end{subequations}
The reformulated dual sub-problem (\ref{eq:trandual}) is a mixed integer second-order cone programming (MISOCP) with the integer variable $\bxi$ and the second-order cone constraint (\ref{eq:socp}).
 Note that the dimension of $\bxi$ is $T$, i.e., the number of time periods, which is independent of the  power network and DERs. Hence, the computational complexity caused by the integer variables does not scale up much with the size of the distribution system.

\begin{figure}[thpb]
	\centering
	\includegraphics[scale=0.24]{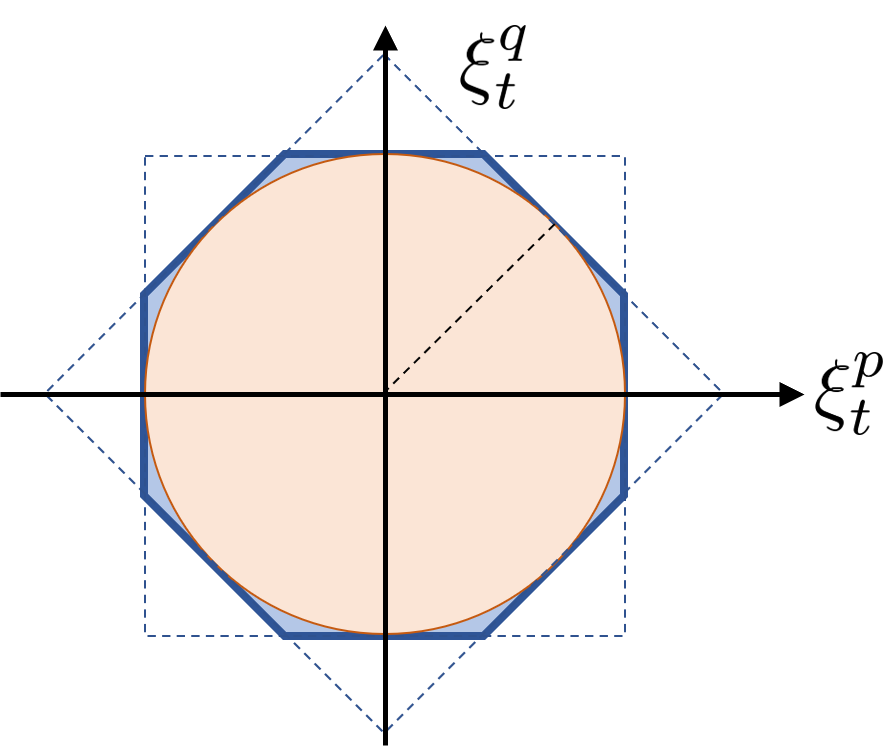}
	\caption{ Linearization method  for the circular quadratic constraint $||\bxi_t||\leq 1$.}	\label{linearization}
\end{figure}

In terms of the sub-problem of the ARPA model (\ref{eq:ARPA}), the uncertainty set $\mathbb{U}_2$ (\ref{eq:ARPA:uncerset}) is a time-decoupling circular region, which has infinite extreme points. Thus the circular constraint linearization method introduced in \cite{rop3} is used to approximate $\mathbb{U}_2$ with the polyhedral uncertainty set $\hat{\mathbb{U}}_2$:
\begin{align*}\label{eq:U2hat}
\begin{split}
      &  \hat{\mathbb{U}}_2: = \!\left\{\bxi_t \!=\! (\xi_t^p,\xi_t^q)\, |  \!-\!1\leq \xi_t^p\leq \!1, \, \!-\sqrt{2}  \leq   \xi_t^p+  \xi_t^q\leq \!\sqrt{2}, \right. \\ &\qquad\qquad \left.
            -1\leq \xi_t^q\leq 1,\,    -\sqrt{2}   \leq  \xi_t^p- \xi_t^q\leq \!\sqrt{2}
    ,\ \forall t\in \mathcal{T}\right\}.
\end{split}
\end{align*}
As illustrated in Figure \ref{linearization}, the approximation uses
two square constraints with the rotation angle of 45 degree to replace the circular region, and
the usage of more  square constraints can enhance the linearization accuracy. Since
$\hat{\mathbb{U}}_2$ is an outer  approximation of ${\mathbb{U}}_2$,  it preserves the robustness of solutions. In other words, the first-stage solutions generated with $\hat{\mathbb{U}}_2$ must be feasible to the original problem with the uncertainty set $\mathbb{U}_2$. Then the optimal  $\bxi_t$ can be parameterized by 
\begin{align*}
\quad    \bxi_t = \sum_{i=1}^n z_{i,t} \bm{e}_i,\ z_{i,t}\in\{0,1\}, \ \sum_{i=1}^n z_{i,t}=1, \ \forall t\in\mathcal{T}
\end{align*}
where $\bm{e}_i\in \mathbb{R}^2$ is one of the  extreme points of the approximate polyhedron, and $n=8$ for the case of $\hat{\mathbb{U}}_2$. As a result,  the same reformulation technique above can be applied to obtain  a tractable optimization model for the  sub-problem of ARPA model (\ref{eq:ARPA}), which is also a MISOCP.

\subsection{Column and Constraint Generation Algorithm}

Based on the master-sub decomposition, the CCG algorithm \cite{ccg} for solving the APA model (\ref{eq:APA}) is presented as Algorithm \ref{ccgalg}. 
%The master problems of the ARA model and ARPA model are LP and SDP respectively, and their corresponding sub-problems are both MILP.  These optimization problems 
 According to \cite[Proposition 2]{ccg}, the CCG algorithm is guaranteed to generate the optimal solution  within a finite number of iterations in the order of $\mathcal{O}(L)$, where $L$ is the number of  extreme points in the uncertainty set. Besides, the master problem and sub-problem can be  solved efficiently with many available optimizers, such as IBM CPLEX and Gurobi.
%{\color{red} I roughly understand the algorithm framework. I think more discussion should be provided: how many iterations would be needed? Can we guarantee that the alg will terminate? Provide some performance discussion, both worst case performance and average performance, in the section. I think you can reduce the some space in Section II so you have space for the performance discussion.}
 
 \begin{algorithm}
 \caption{Column and Constraint Generation Algorithm}
 \begin{algorithmic}[1] \label{ccgalg}
 
   \STATE \textbf{Initialization: } Set $K=1$  and tolerance $\epsilon> 0$. 
    Initialize $\bxi$ with an appropriate  value, e.g., 
     $\bxi^1_* = \bm{1}$ or  $\bxi^1_* = \bm{0}$, and  $f_S$ with a large value.
     
   \WHILE{$|f_S| \geq \epsilon$} 

      \STATE  	- \textbf{Solve Master Problem} (\ref{eq:master}) to obtain the optimal power aggregation solution  $(\bp_{0*}^{\vee},\bp_{0*}^{\wedge})$.
      \STATE    	- \textbf{Solve Sub Problem} (\ref{eq:trandual}) with given  $(\bp_{0*}^{\vee},\bp_{0*}^{\wedge})$ to obtain the optimal $\bxi_*^{K+1}$ and the objective value $f_S$. 
      
      \STATE - Generate new variables $\bx^{K+1}$  and add new constraints (\ref{eq:newcon}) to the master problem (\ref{eq:master}).
	\begin{subequations} \label{eq:newcon}
	    	\begin{align}
        &   \bm{\xi}^{K+1}_* \circ \bp^\wedge_0 + (\bm{1} -\bm{\xi}^{K+1}_*)\circ\bp^\vee_0  = \mb{D}\bx^{K+1} + \bm{g}\\
       & ||\mb{E}_l \bx^{K+1}||\leq {s}_l, \ 
    \forall l\in\mathcal{L}\\
        & \mb{W}\bx^{K+1} \leq \bm{w}
	\end{align}
	\end{subequations}
	$\ $ Update $K\leftarrow K+1$. 
    \ENDWHILE     
    \STATE \textbf{Output} the final feasible intervals $(\bp_{0*}^{\vee},\bp_{0*}^{\wedge})$.
 \end{algorithmic} 
 \end{algorithm}

\begin{figure}
    \centering
    \includegraphics[scale=0.57]{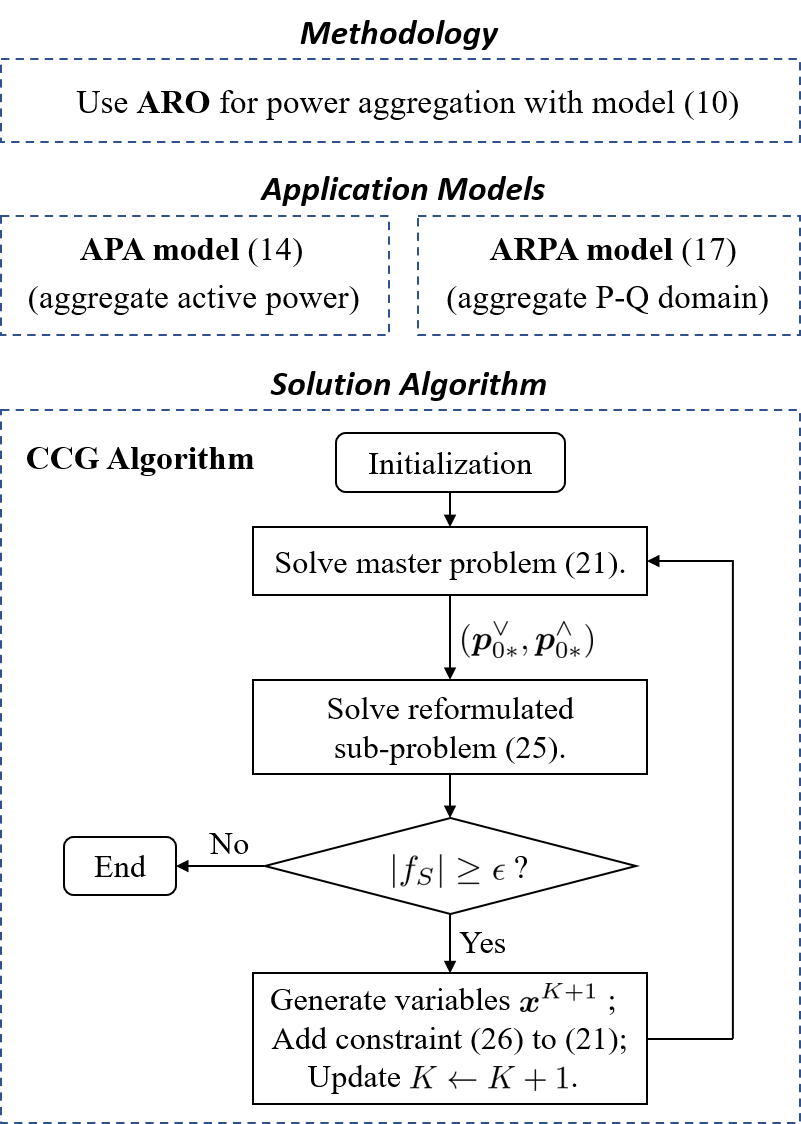}
    \caption{The structure of the proposed power flexibility aggregation method and the solution flowchart.} 
    \label{fig:structure}
\end{figure}

The overall structure of the proposed power  aggregation method and the solution algorithm is illustrated as Figure \ref{fig:structure}. 

%% file: case_study.tex
\section{Numerical Simulation}

\begin{figure}[thpb]
	\centering
	\includegraphics[scale=0.32]{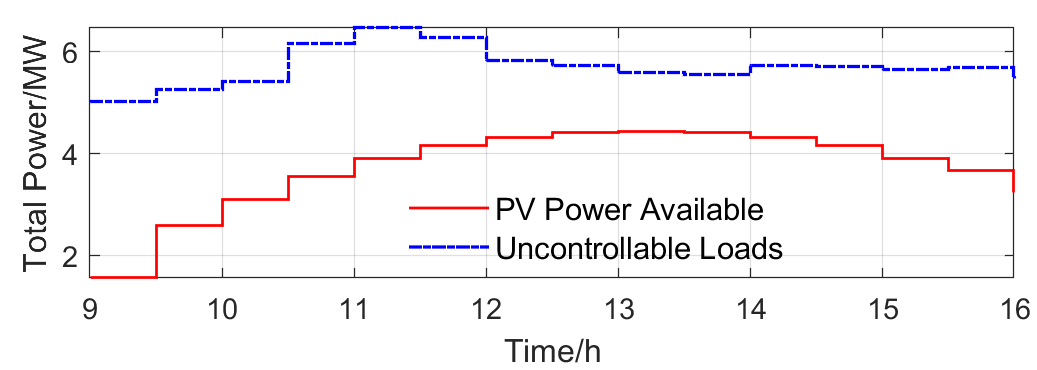}%{L_PV}
	\caption{Total PV power available and uncontrollable loads from 9:00 to 16:00 with the granularity of 30 minutes.}
	\label{fig:L_PV}
\end{figure}

Numerical tests are carried out on a real unbalanced distribution feeder located within the territory of  Southern California Edison (SCE). This feeder contains 126 multi-phase buses with a total of 366 single-phase connections. The nominal voltage at the substation is 12kV (1 p.u.), and we set the upper and lower limits of voltage magnitude as 1.02 p.u. and 0.98 p.u. Dispatchable DERs include 33 PV units, 28 ES devices and 5 HVAC systems. The real load data and real solar irradiance profiles   are applied. The total amounts of uncontrollable loads and available PV power from 9:00 to 16:00 are presented as Figure \ref{fig:L_PV}. For PV units, we set   the lower bound of power generation as zero and take the available PV power in Figure \ref{fig:L_PV} as the upper bound.
 We set the initial SOC of the ES devices to $50\%$ and the storage efficiency factor $\kappa_i$ to $0.95$. The simulation time is discretized with the granularity of 30 minutes. Detailed configurations and parameters of this feeder system are provided in \cite{sys}.

\subsection{Implementation of Active Power Aggregation} \label{sec:sim:com}

\begin{figure}[thpb]
	\centering
	\includegraphics[scale=0.32]{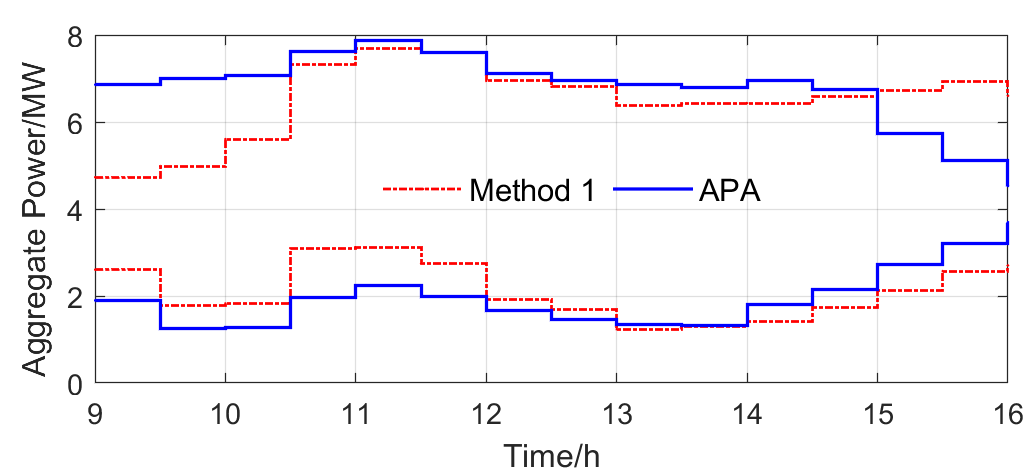}%{L_PV}
	\caption{The optimal feasible intervals $[p_{0,t}^{\vee*}, p_{0,t}^{\wedge*}]_{\,t\in\mathcal{T}}$ of aggregate active power from 9:00 to 16:00 obtained by APA model and Method 1.}
	\label{fig:apa_pre}
\end{figure}

We implemented the APA model (\ref{eq:APA}) to evaluate the
maximal active power flexibility of the test system, and compared the results with our previous method in \cite{our1} (Method 1). The feasible intervals of aggregate active power obtained via the APA model and Method 1 are illustrated as Figure \ref{fig:apa_pre}. Define the aggregate flexibility as $E_{af} = \sum_{t\in\mathcal{T}} (p_{0,t}^\wedge - p_{0,t}^\vee)\cdot\Delta t$. Then the aggregate flexibility values associated with the APA model and Method 1 are $35.39$ MWh and $32.90$ MWh respectively, i.e., $2.49$ MWh more flexibility can be extracted by using the APA scheme. That is because Method 1 imposes conservative  constraints (\ref{eq:heu}) on  ES power and HVAC power to ensure  disaggregation feasibility, while the APA scheme guarantees this property with optimality by leveraging the ARO modelling technique. Besides, we tuned the total ES capacity in the test system, and 
 compared the performance of the two methods above. 
The aggregate flexibility obtained via the APA model and Method 1 is shown in Table \ref{tab:com_bat}. The results further validate that Method 1 does not fully exploit the ES power flexibility due to the conservative constraints,
and the superiority of the APA scheme is more significant in the case with  higher  penetration of ES (or HVAC) facilities.

\begin{table}[!t]
\renewcommand{\arraystretch}{1.3}
\caption{The aggregate flexibility comparison between APA and Method 1 with different ES capacities.}
\label{tab:com_bat}
\centering
\begin{tabular}{cccccc}
\hline\hline
  \multicolumn{2}{c}{Total ES capacity/MWh}
  & 2.72
 & 5.44
 &  8.17 & 10.89
 \\  \hline
 \multirow{2}{*}{$E_{af}$/MWh} & APA   & {33.59} & {35.39} &  {36.74} & {38.00}	\\ 
& Method 1 & {32.74}  &  {32.90} &   { 33.04}   & {33.13}\\ \hline \hline
\end{tabular}
\end{table}

\subsection{Implementation of Power Disaggregation}

\begin{figure}[thpb]
	\centering
	\includegraphics[scale=0.32]{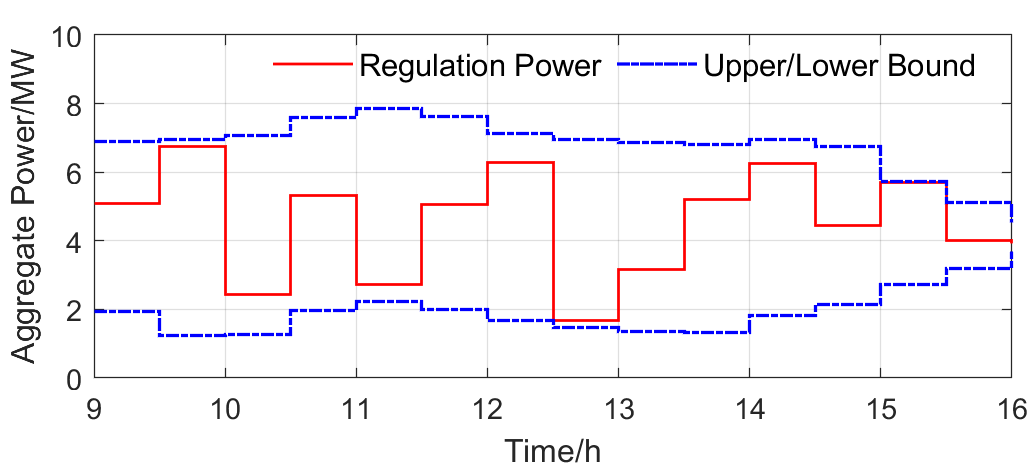}%{L_PV}
	\caption{The optimal feasible interval $[p_{0,t}^{\vee*}, p_{0,t}^{\wedge*}]_{\,t\in\mathcal{T}} $ (in blue) of aggregate active power and the randomly generated regulation power trajectory (in red).}
	\label{fig:aggP}
\end{figure}

\begin{figure}[thpb]
	\centering
	\includegraphics[scale=0.32]{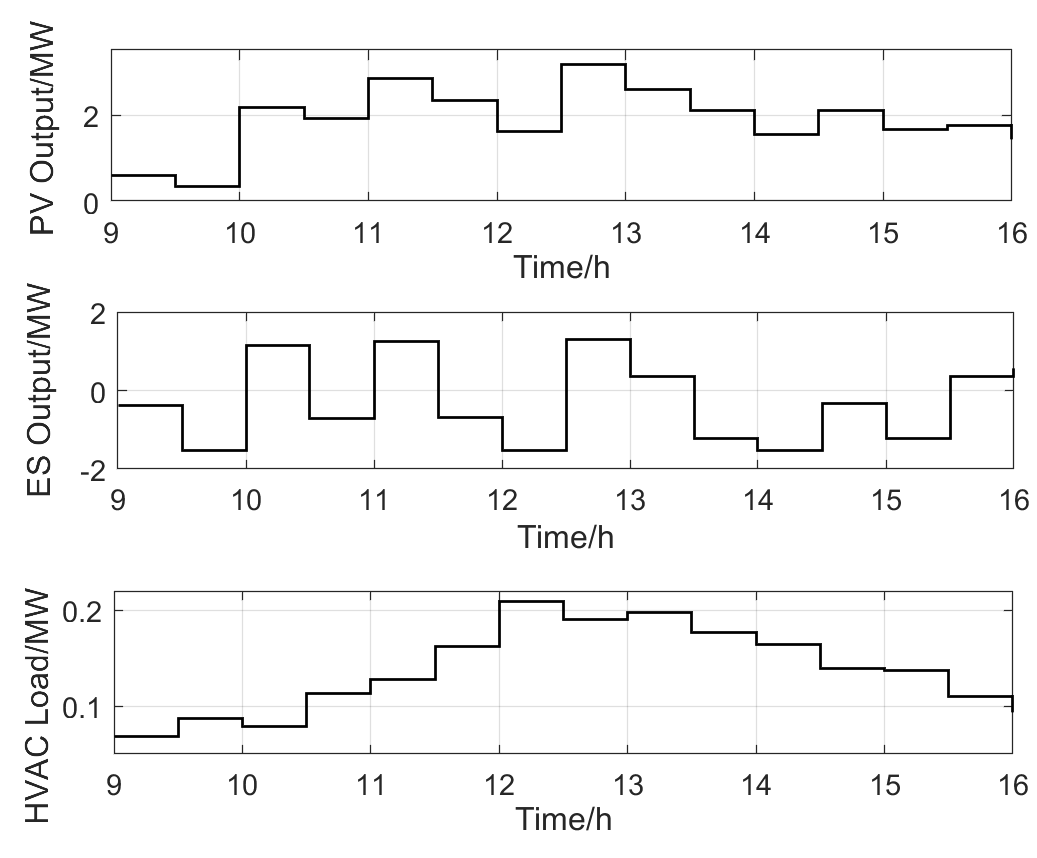}%{L_PV}
	\caption{The optimal DER power dispatch scheme $\bx^*$ to track the regulation power trajectory in Figure \ref{fig:aggP}.}
	\label{fig:DER}
\end{figure}

We performed the Monte Carlo simulation to 
verify the disaggregation feasibility of the  feasible intervals $[p_{0,t}^{\vee*}, p_{0,t}^{\wedge*}]_{\,t\in\mathcal{T}} $ obtained by the APA model in Figure \ref{fig:apa_pre}. We  assume that the regulation power trajectory $ (p_{0,t}^{\mathrm{reg}})_{t\in \mathcal{T}}$ from the transmission-level dispatch are  random variables following the uniform  distribution, i.e., $p_{0,t}^{\mathrm{reg}} \sim \text{Unif}\left(p_{0,t}^{\vee*}, p_{0,t}^{\wedge*} \right)$ independently for each $t\in \mathcal{T}$. 
Up to 3000  regulation power trajectories are  randomly generated, and we solve the PD problem (\ref{eq:PD})  for each case.
The simulation results show that the PD problem (\ref{eq:PD}) is feasible for all the generated power trajectories $(p_{0,t}^{\mathrm{reg}})_{t\in \mathcal{T}}$, and 
  we can always obtain a corresponding optimal DER dispatch scheme $\bx^*$ for each of them.  One of the cases is presented as Figure \ref{fig:aggP} and Figure \ref{fig:DER}. The PV output, ES output, and HVAC load in Figure \ref{fig:DER} denote the summed values over the same type of DERs.  These results numerically validate the  disaggregation feasibility of the proposed  method.
  

\subsection{Implementation of Active-Reactive Power Aggregation}

\begin{figure}[thpb]
	\centering
	\includegraphics[scale=0.29]{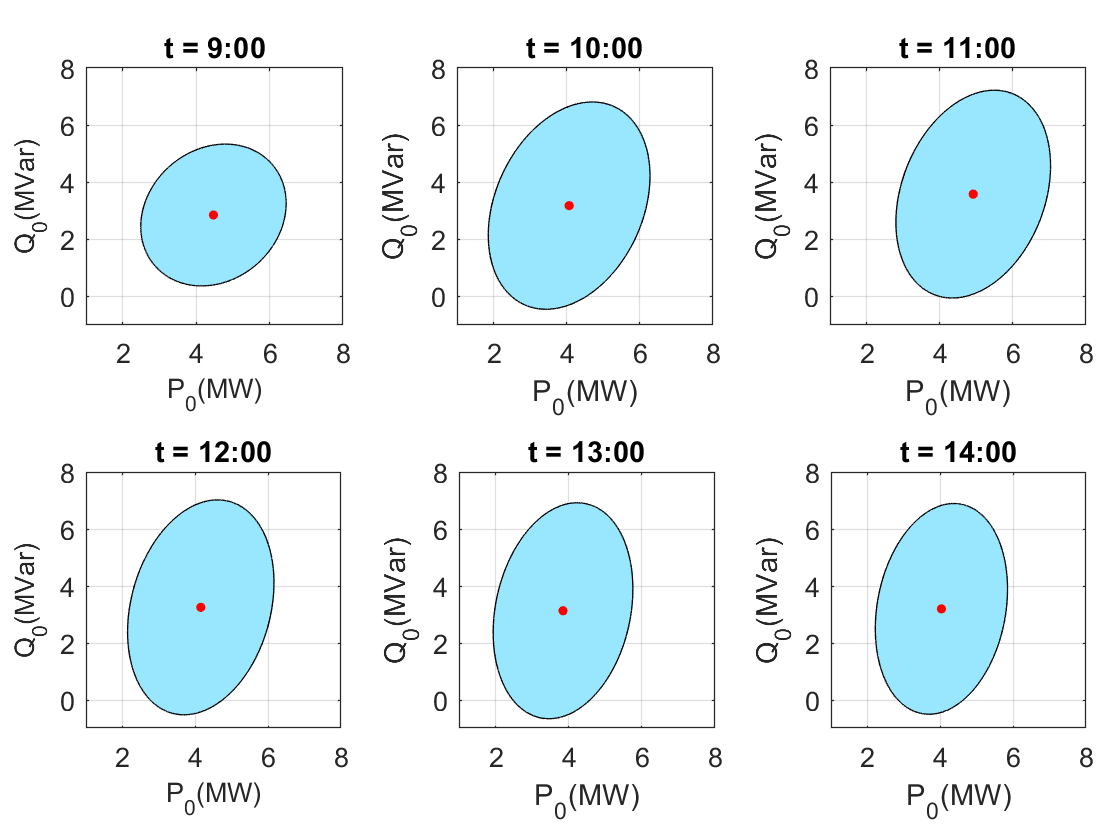}%{L_PV}
	\caption{The time-variant elliptical feasible regions for the aggregate active-reactive power domain from 9:00 to 14:00.}
	\label{fig:P-Q_2}
\end{figure}

We  implemented the ARPA model (\ref{eq:ARPA}) to compute the elliptical feasible regions for the aggregate P-Q domain. 
The simulation time horizon is selected from 9:00 to 14:00  with the granularity of 1 hour. 
%To explicitly exhibit the relation between the aggregate flexibility and available DER capacity, we removed the ES devices, and only considered PV units and HVAC systems. 
The P-Q flexibility aggregation results are illustrated as Figure \ref{fig:P-Q_2}, where the red dot is the center $(p_{0,t}^c, q_{0,t}^c)$ and the blue areas represent the feasible regions of aggregate P-Q over time.

%Since the PV unitsvaccount for most aggregate flexibility,  it is observed that the size and shape of the elliptical feasible regions are highly related to the PV available power and uncontrollable loads shown in Figure \ref{fig:L_PV}. 

%\begin{figure}[thpb]
%	\centering
%	\includegraphics[scale=0.35]{P_Q_no_bat.png}%{L_PV}
%	\caption{The time-variant elliptical feasible regions for the aggregate active-reactive power domain from 9:00 to 14:00.}
%	\label{fig:P-Q}
%\end{figure}

\begin{figure}[thpb]
	\centering
	\includegraphics[scale=0.32]{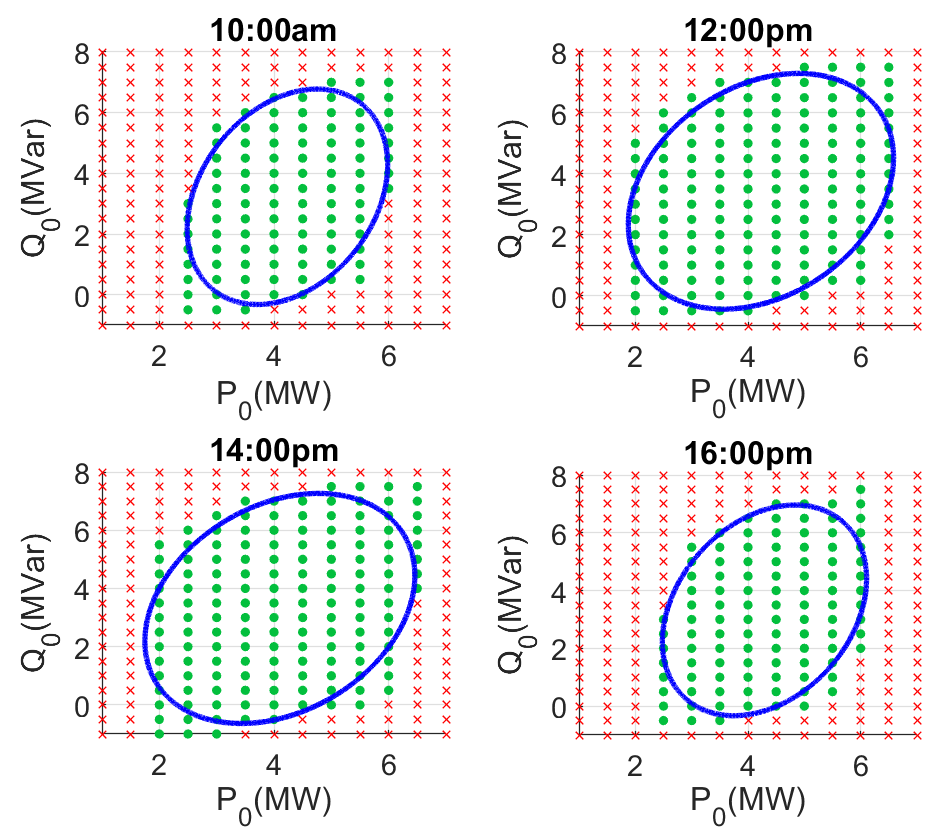}%{L_PV}
	\caption{Test of disaggregation feasibility and the
	elliptical feasible regions  computed via ARPA. (Red crosses denote infeasible  aggregate P-Q points; green dots denote feasible  aggregate P-Q points; blue curves denote the computed elliptical feasible regions of aggregate P-Q power.)}
	\label{fig:PQcom}
\end{figure}

Furthermore, we quantized the aggregate P-Q domain with the resolution of $0.5$ MW(MVar), and tested the disaggregation feasibility of every single point with corresponding $(\bm{p}_0,\bm{q}_0)$ by solving the power disaggregation problem (\ref{eq:pqpd}).
\begin{align} \label{eq:pqpd}
  \text{Obj.}  \,    \min_{\bx} 0\qquad
  \text{s.t.} \ \text{Equation (\ref{fin})}.
\end{align}
The results are  compared with the  elliptical feasible regions computed by the ARPA model. To make better comparison, we considered a single time step with $T=1$ to avoid the time-coupling impact, and ran the simulations at four different times separately, i.e., $10$am, $12$pm, $14$pm and $16$pm. The simulation results are shown as Figure \ref{fig:PQcom}. The areas consisting of green dots are essentially the actual feasible regions of aggregate P-Q.  It is observed that the obtained	elliptical  regions lie within the green areas in all cases, which validates the disaggregation feasibility of the proposed method. Moreover, the elliptical  region covers most of the green area, which indicates that the proposed method can obtain an accurate 	elliptical approximation of the actual feasible region. Besides,  the feasible regions at $12$pm and $14$pm are larger than those at $10$am and $16$pm. This is mainly because the available PV generation is higher at  $12$pm and $14$pm, leading to more power flexibility. Note that the 
	elliptical feasible regions in Figure \ref{fig:PQcom} are not necessarily the same as those at the same time  in Figure \ref{fig:P-Q_2}, since the 
	elliptical feasible regions in Figure \ref{fig:P-Q_2} are computed by considering a holistic time horizon from 9am to 14pm.

\subsection{Computational Efficiency}

Simulations are performed in a computing environment with Intel(R) Core(TM) i7-7660U CPUs running at 2.50 GHz and with 8-GB RAM. 
All the programmings are implemented in Matlab 2018b. We use the CVX package \cite{cvx1} to model the convex programs,  solve SDP with SDPT3 solver \cite{SDPT}, and solve mixed integer programs with  Gurobi  optimizer \cite{gur}.

For the simulations above, 
 the average solution times for the master problem (\ref{eq:master}) and the sub-problem (\ref{eq:trandual}) of the APA model (\ref{eq:APA}) are 9.25s and 213.9s, respectively. With the initial uncertainty scenarios $\bxi^1_* = \mathbf{1}$ and $\bxi^2_* = \mathbf{0}$, the CCG algorithm usually converges within one or two iterations. In terms of the 
ARPA model (\ref{eq:ARPA}),
it takes 145.2s and 571.6s on average to solve   the corresponding master  problem   and  sub-problem.  
Since power flexibility aggregation is a hours-ahead/day-ahead scheduling problem, the computational time is generally acceptable for practical application. Besides, it is  observed that  much more time is spent in solving the sub-problems, because the sub-problems are in the form of nonconvex mixed integer programs.  To accelerate the computation,   some approximation methods, such as 
the outer approximation algorithm introduced in \cite{oa}, can be employed to address the bilinear issue in the sub-problems instead of modelling with integer variables.

%% file: conclusion.tex
% ------------------------------------------
\section{Conclusion}

In this paper, based on the ARO framework, we propose a novel methodology to perform power flexibility aggregation for unbalanced distribution systems with heterogeneous DER facilities. Two concrete aggregation models (APA and ARPA) with two-stage optimization are developed for  different implementation goals. 
The APA model aggregates purely active power and computes its optimal feasible intervals over time, while the ARPA model characterizes both active and reactive power flexibility and solves the time-variant elliptical feasible regions of the aggregate P-Q domain. Lastly, the numerical tests on a real distribution feeder validate that both the aggregation optimality and disaggregation feasibility are guaranteed with the proposed method. 
  Future works include 1) coordinating the power flexibility aggregation with the real-time DER control to exploit additional flexibility and facilitate practical application; 2)  developing distributed solution algorithm for the proposed aggregation models to enhance the computational efficiency and preserve the privacy of DERs.

%Future work is to develop time-coupling formulations of the aggregate feasible region to fully exploit the DER flexibility.

%% file: appendix.tex
\appendices

\section{Linear Multi-Phase Power Flow Model} \label{lpfm}
%\lina{notations are unclear; what the $\hat{v},\hat{x}_Y$ etc, don't use $\dot{v}$; conjugate can use $V^*$}

For expression simplification, we discard the time subscript $t$ in the following notations.  Let column vectors $\bm{s}_Y:=\bm{p}_Y+\jmath\,\bm{q}_Y$ and $\bm{s}_\Delta:=\bm{p}_\Delta+\jmath\,\bm{q}_\Delta$ collect all the three-phase complex power injection via wye- and delta-connection, respectively. Define  vectors
 $\bm{x}_Y:=\left[\bm{p}_Y^\top,\bm{q}_Y^\top \right]^\top$, $\bm{x}_\Delta:=\left[\bm{p}_\Delta^\top,\bm{q}_\Delta^\top \right]^\top$.
 
Let $\tilde{\bv}\in \mathbb{C}^{3|\mathcal{N}|}$ be the column vector collecting all the three-phase nodal complex voltage. Reference \cite{linear_1} shows that the complex voltage vector $\tilde{\bv}$ satisfies the following fixed-point equation (\ref{com_v}): 
\begin{align} \label{com_v}
\tilde{\bv}= \mb{M}_Y(\tilde{\bv})\cdot\bm{x}_Y + \mb{M}_\Delta(\tilde{\bv}) \cdot\bm{x}_\Delta + \bm{m}.
\end{align}
See \cite{linear_1} for the detailed definitions of the matrix functions $\mb{M}_Y(\tilde{\bv}), \mb{M}_\Delta(\tilde{\bv})$ and parameter $\bm{m}$.
Based on a given operational point $\{\tilde{\bv}^{\text{o}},\bx_Y^{\text{o}},\bx_\Delta^{\text{o}} \}$, it aims to derive a linear approximate model for the voltage magnitudes $\bv \!:= \!|\tilde{\bv}|$. To this end, we leverage the following derivation rule
\begin{align*}
\frac{\partial |f(x)|}{\partial x} = \frac{1}{|f(x)|} \mathcal{R} \left\{ f(x)^* \frac{\partial f(x)}{\partial x}  \right\},
\end{align*}
and obtain the linear model (\ref{eq:voltfor}) based on equation (\ref{com_v}).
\begin{align} \label{eq:voltfor}
    {\bv}= \mb{A}_Y\cdot\bm{x}_Y + \mb{A}_\Delta \cdot\bm{x}_\Delta + \bm{a}.
\end{align}
where 	$\mathcal{R}\{\cdot\}$ denotes the real part of a complex value, $(\cdot)^*$ denotes the complex conjugate, and
\begin{subequations}
\begin{align}
\mb{A}_Y&:= \frac{\partial |\tilde{\bv}|}{\partial \bx_Y}= \text{diag}(|{\tilde{{\bv}}}^{\text{o}}|)^{-1}\mathcal{R}\left\{ \text{diag}({\tilde{{\bv}}}^{\text{o}*})\mb{M}_Y(\tilde{\bv}^{\text{o}})  \right\}\\
\mb{A}_\Delta&:=\frac{\partial |\tilde{\bv}|}{\partial \bx_\Delta}=\text{diag}(|{\tilde{{\bv}}}^{{\text{o}}}|)^{-1}\mathcal{R}\left\{ \text{diag}({\tilde{{\bv}}}^{{\text{o}}*})\mb{M}_\Delta (\tilde{\bv}^{\text{o}}) \right\}\\
\bm{a}&:=|{\tilde{{\bv}}}^{\text{o}}|-\mb{A}_Y{\bm{x}}_Y^{\text{o}}-\mb{A}_\Delta{\bm{x}}_\Delta^{\text{o}} \label{eq:a}.
\end{align}
\end{subequations}
Essentially, the linear model (\ref{eq:voltfor}) can be viewed as a
linear interpolation between two power flow points: the given operational point $\{\tilde{\bv}^{\text{o}},\bx_Y^{\text{o}},\bx_\Delta^{\text{o}} \}$ and the operational point with zero power injection. Define $\bm{x}:=\left[{\bm{x}_Y}^\top, {\bm{x}_\Delta}^\top \right]^\top$. Then we obtain $\mb{A}:=\left[\mb{A}_Y,\mb{A}_\Delta \right]$ in (\ref{v}), and $\bm{a}_t$ in  (\ref{v}) is the counterpart of $\bm{a}$ (\ref{eq:a}) at time $t$.

Similarly, 
using Kirchhoff's laws, we can further derive matrix  $\mb{B}$, vectors $\bm{b}_t,\bm{d},\bm{f}$ and scalars $g_t,h_t$ in (\ref{lp1}). 
See reference \cite{linear_2} for  details.

\section{Derivation of Dual Sub-Problem} \label{app:duality}

For (\ref{eq:sub:qua}), we define a new variable
 $\by_l = \mb{E}_l \bx$ and transform it to $||\by_l||\leq {s}_l$ for all $l\in\mathcal{L}$. Then the Lagrangian function of the sub-problem (\ref{eq:sub}) is formulated as (\ref{eq:lagr}) with dual variables $\bmu,\bga_l,\sigma_l\geq 0, \bm{\lambda}\geq \bm{0}$.
 \begin{align} \label{eq:lagr}
     L& =  \big(  \bm{\xi} \circ (\bm{p}^\wedge_{0*}- \bm{p}^\vee_{0*})+ \bm{p}^\vee_{0*}-\mb{D}\bx - \bm{g} \big)^\top \bmu \nonumber\\
     & + \sum_{l\in\mathcal{L}}( (\by_l- \mb{E}_l \bx)^\top \bga_l + \sigma_l(s_l-||\by_l||) ) +  (\bm{w}-\mb{W}\bx)^\top \bm{\lambda} \nonumber\\
     & = \underbrace{(\bm{p}^\wedge_{0*}- \bm{p}^\vee_{0*})^\top(\bmu\circ\bxi) +\!  (\bm{p}^\vee_{0*}\!- \bm{g} )^\top \bmu +\! \sum_{l\in\mathcal{L}} s_l\sigma_l +\bm{w}^\top \bm{\lambda}}_{:=\mathcal{H}(\bxi,\bmu,(\sigma_l)_{l\in\mathcal{L}},\bm{\lambda})} \nonumber\\
     & + \bx^\top\big(- \mb{D}^\top \bmu  - \sum_{l\in\mathcal{L}} \mb{E}_l^\top \bga_l - \mb{W}^\top\bm{\lambda}  \big) \nonumber\\
     &+ \sum_{l\in\mathcal{L}} \big(\bga_l^\top\by_l- \sigma_l||\by_l||   \big) 
 \end{align}
To maximize (\ref{eq:lagr}) over $\bx$ and $(\by_l)_{l\in\mathcal{L}}$, we have 
\begin{subequations}\label{eq:dualcondi}
\begin{align}
    &\mb{D}^\top \bmu  + \sum_{l\in\mathcal{L}} \mb{E}_l^\top \bga_l + \mb{W}^\top\bm{\lambda} = \bm{0} \\
    & ||\bga_l||\leq \sigma_l,\quad \forall l\in\mathcal{L}\label{eq:dualcondi:socp}
\end{align}
\end{subequations}
and the result (\ref{eq:finalres}), which leads to the dual form (\ref{eq:dual}). Note that $\sigma_l\geq 0$
is directly guaranteed with  constraint (\ref{eq:dualcondi:socp}).
\begin{align} \label{eq:finalres}
     \max_{\bx,\by_l} L  =\begin{cases}
     \mathcal{H}(\bxi,\bmu,(\sigma_l)_{l\in\mathcal{L}},\bm{\lambda}), & \text{if (\ref{eq:dualcondi}) is satisfied.}\\
     +\infty, & \text{otherwise.}
     \end{cases}
 \end{align}

%% file: biography.tex

\vskip -4pt plus -1fil

\begin{IEEEbiography}[{\includegraphics[width=1in,height=1.25in,clip,keepaspectratio]{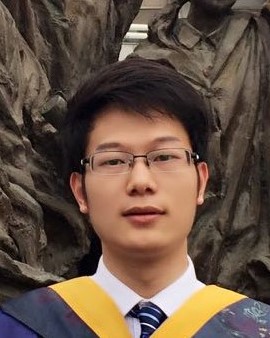}}]{Xin Chen}
received his double B.S. degrees in engineering physics and economics from Tsinghua University, Beijing, China in 2015, and the master degree in electrical engineering from Tsinghua University, Beijing, China in 2017. He is 
currently pursuing the Ph.D. degree in electrical engineering at Harvard University, USA. 
He received the Best Conference Paper Award in IEEE PES General Meeting 2016, and he was a Best Student Paper Award Finalist in the IEEE Conference on Control Technology and Applications (CCTA) 2018. His research interests focus on 
distributed optimization, control, and learning of networked systems, with emphasis on power systems.
\end{IEEEbiography}

\vskip 0pt plus -1fil

\begin{IEEEbiography}[{\includegraphics[width=1in,height=1.25in,clip,keepaspectratio]{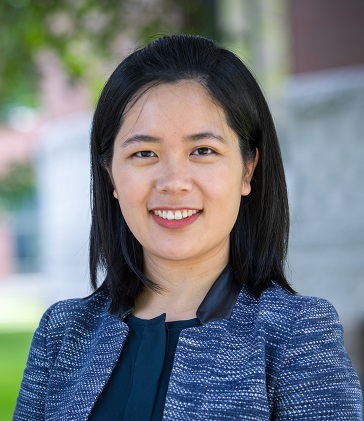}}]{Na Li}
 received her B.S. degree in mathematics and applied mathematics from Zhejiang University in
	China and her PhD degree in Control and Dynamical Systems from the California Institute of Technology
	in 2013. She is an Associate Professor in the School of Engineering and Applied Sciences at Harvard
	University. She was a postdoctoral associate of the Laboratory for Information and Decision Systems at
	Massachusetts Institute of Technology. Her research lies in the design, analysis, optimization,
	and control of distributed network systems, with particular applications to cyber-physical network systems. She received NSF CAREER Award in 2016, AFOSR Young Investigator Award in 2017,
	ONR Young Investigator Award in 2019 among others. 
\end{IEEEbiography}
